\documentclass[12pt]{article} 

\usepackage{geometry} 
\usepackage{graphicx} 
\usepackage{lineno,hyperref}
\usepackage{graphicx}
\usepackage{rotating}
\usepackage{pdflscape}
\usepackage{amsmath}
\usepackage{epstopdf}
\usepackage{multirow}
\usepackage{bbm}
\usepackage{bm}
\usepackage{algorithm}
\usepackage[noend]{algpseudocode}
\usepackage{charter}
\usepackage{color}
\usepackage[round]{natbib}

\newcommand{\indep}{\;\, \rule[0em]{.03em}{.6em}\hspace{-.25em}\rule[0em]{.65em}{.03em}\hspace{-.25em}\rule[0em]{.03em}{.6em}\;\,}
\modulolinenumbers[5]

\newcommand{\argmaxF}{\mathop{\mathrm{argmax}}} 

\usepackage{booktabs} 
\usepackage{array} 
\usepackage{paralist} 
\usepackage{verbatim} 
\usepackage{subfig} 

\usepackage{fancyhdr} 
\usepackage{sectsty}
\allsectionsfont{\sffamily\mdseries\upshape} 

\usepackage[nottoc,notlof,notlot]{tocbibind} 
\usepackage[titles,subfigure]{tocloft} 

\title{New Parsimonious Multivariate Spatial Model: Spatial Envelope }
\author{
Hossein Moradi Rekabdarkolaee$\!^a$, Qin Wang$\!^a$, Zahra Naji, Montserrat Fuentes$\!^a$ \\[0.2in] 
\normalsize\itshape $^a$ Department of Statistical Sciences and Operations Research,\\ 
\normalsize\itshape Virginia Commonwealth University, Richmond, VA 23284, USA\\[0.05in] 
}
\date{} 

\begin{document}
\maketitle

\hrule
\begin{abstract}
Dimension reduction provides a useful tool for analyzing high dimensional data. The recently developed \textit{Envelope} method is a parsimonious version of the classical multivariate regression model through identifying a minimal reducing subspace of the responses. However, existing envelope methods assume an independent error structure in the model. While the assumption of independence is convenient, it does not address the additional complications associated with spatial or temporal correlations in the data. In this article, we introduce a \textit{Spatial Envelope} method for dimension reduction in the presence of dependencies across space. We study the asymptotic properties of the proposed estimators and show that the asymptotic variance of the estimated regression coefficients under the spatial envelope model is smaller than that from the traditional maximum likelihood estimation. Furthermore, we present a computationally efficient approach for inference. The efficacy of the new approach is investigated through simulation studies and an analysis of an Air Quality Standard (AQS) dataset from the Environmental Protection Agency (EPA).
\end{abstract}




\vspace{0.1in}
\hrule


\section{Introduction}

In many research areas, such as  health science \citep{lave1973analysis,liang1992multivariate}, environmental sciences \citep{guinness2014multivariate}, and business \citep{cooper2003business}, etc., it is common to observe multiple outcomes simultaneously. The traditional multivariate linear model has proved to be useful in these cases to understand the relationship between response variables and predictors. Mathematically, the model is typically presented as:
\begin{equation}\label{en111}
\textbf{Y}=\boldsymbol\alpha+\boldsymbol\beta \textbf{X} +\boldsymbol\epsilon,
\end{equation}
where $\textbf{Y} \in \mathbbm{R}^r$ denotes the response vector, $\textbf{X}\in \mathbbm{R}^p$ is a predictor vector, $\boldsymbol\alpha \in \mathbbm{R}^{r}$ denotes the vector of intercept, $\boldsymbol\beta \in \mathbbm{R}^{(r\times p)}$ is the matrix of regression coefficients, and $\boldsymbol\epsilon \sim N_{r}(\textbf{0},\boldsymbol\Sigma)$ is an error vector with $\boldsymbol\Sigma \geq 0$ being an unknown covariance matrix  \citep{christensen2001advanced}. In order to completely specify a multivariate linear model, there are $r$ unknown intercepts, $p\times r$  unknown parameters for the matrix of regression coefficients, and $r(r+1)/2$ unknown parameters  to specify an unstructured covariance matrix. Therefore, one must estimate $r+pr+r(r+1)/2$ parameters which can be large with the increase of either $r$ or $p$ or both.

Based on the observation that some linear combinations of \textbf{Y} do not depend on any of the predictors in some cases, \cite{cook2010envelope} proposed the \textit{Envelope} method as a parsimonious version of the classical multivariate linear model. This approach separates the \textbf{Y} into material and immaterial parts, thereby allowing gains in estimation efficiency compared to the usual maximum likelihood estimation. The envelope approach constructs a link between the mean function and covariance matrix using a minimal reducing subspace such that the resulting number of parameters will be maximally reduced. \cite{cook2010envelope} showed that the envelope estimator are at least as efficient as the standard maximum likelihood estimator (MLE). Along the same line, the idea of envelope has been further developed from both theoretical and computational points of view in a series of papers including, but not restricted to, \cite{su2011partial, su2012inner, su2013estimation}, \cite{cook2015simultaneous}, and \cite{cook2016note}. Furthermore, \cite{li2017parsimonious} and \cite{zhang2017tensor} extended the envelope model to the tensor response and tensor coviariates, respectively. 

Proposed envelope methodology by \cite{cook2010envelope} assumes observations are taken under identical conditions where independence is assured. While models based on the independence assumption are extremely useful, their use is limited in applications where the data has inherent dependency \citep{cressie2015statistics}. For example, in environment monitoring, each station collects data concerning several pollutants such as ozone, carbon monoxide, nitrogen dioxide, etc. These data have a special type of dependency which is called spatial correlation. \cite{myers1991pseudo} and \cite{ver1998constructing} used pseudo cross-variogram to model the multivariate spatial cross-correlation. In addition, \cite{chiles2009geostatistics} and \cite{wackernagel2013multivariate} introduced several multivariate covariogram and cross-variogram that results in a nonnegative definite covariance matrix (also called valid spatial covariance function). Linear Coregionalization Models (LCM) is one the most commonly used approaches in the multivariate spatial data analysis. This model assumes that 
the observed variables are linear combinations of sets of independent underlying variables and they covary jointly over a region. Different methods have been proposed for fitting LCM in literatures including, but not restricted to, least square approach \citep{goulard1992linear}, expectation-maximization (EM) algorithm \citep{zhang2007maximum}, etc. \cite{gneiting2010matern} introduced a flexible and interpretable Matern cross-covariance function for multivariate spatial random field. \cite{genton2015cross} provided a comprehensive review on common approaches for building a valid spatial cross-covariance models. In this paper, we introduce a \textit{Spatial Envelope} approach for spatially correlated data. This new  approach addresses the impact of spatial correlation among observations in the model and thus provides more efficient estimators than the traditional multivariate linear model and linear coregionalization model. Accounting for the intrinsic spatial correlation allows the appropriate inference on aforementioned data. 

The rest of the paper is organized as follows: in section 2, we briefly review envelope methodology. The spatial envelope is detailed in Section 3. Section 4 and 5 provides asymptotic variance and prediction properties of the proposed method. Section 6 and 7 contain a simulation study and the analysis of the northeastern United State air pollution data. We conclude the article with a short discussion in Section 8. All technical details are provided in the Appendix. 


\section{Brief Review of envelope}

For model (\ref{en111}), suppose that we can find an orthogonal matrix $(\boldsymbol\Gamma_{1}, \boldsymbol\Gamma_{0}) \in \mathbbm{R}^{r\times r}$
that satisfies the following two conditions: (i) $span(\boldsymbol\beta) \subseteq span(\boldsymbol\Gamma_{1})$, and (ii) $\boldsymbol\Gamma_{1}^{T}\textbf{Y}$ is conditionally independent of $\boldsymbol\Gamma_{0}^T \textbf{Y}$ given \textbf{X}. That is, $\boldsymbol\Gamma_{0}^{T} \textbf{Y}$ is marginally independent of \textbf{X} and conditionally independent of \textbf{X} given $\boldsymbol\Gamma_{1}^{T}\textbf{Y}$. Then, we can rewrite $\boldsymbol\Sigma$ as
\begin{equation}
\label{en2}
\boldsymbol\Sigma=\textbf{P}_{\boldsymbol\Gamma_{1}}\boldsymbol\Sigma \textbf{P}_{\boldsymbol\Gamma_{1}}+\textbf{Q}_{\boldsymbol\Gamma_{1}}\boldsymbol\Sigma \textbf{Q}_{\boldsymbol\Gamma_{1}},
\end{equation}
where $\textbf{P}_{(\cdot)}$ represents an orthogonal projection operator with respect to the standard inner product and  $\textbf{Q}_{(\cdot)}=\textbf{I}_{r}-\textbf{P}_{(\cdot)}$ is the projection onto its complement space. \cite{cook2010envelope} used this idea to construct the unique smallest subspace $span(\boldsymbol\Gamma_{1})$ that satisfies (\ref{en2}) and contains  $span(\boldsymbol\beta)$. In summary, the goal is to find a subspace $span(\boldsymbol\Gamma_{1}) \subseteq  \mathbbm{R}^r$ such that
\begin{subequations}
\begin{align}
\label{bas1}
\textbf{Q}_{\boldsymbol\Gamma_{1}}\textbf{Y} | \textbf{X} \sim \textbf{Q}_{\boldsymbol\Gamma_{1}}\textbf{Y},\\
\label{bas2}
\textbf{Q}_{\boldsymbol\Gamma_{1}}\textbf{Y}\indep \textbf{P}_{\boldsymbol\Gamma_{1}}\textbf{Y} | \textbf{X}.
\end{align}
\end{subequations}
where $\indep$ means statistical independence. This minimal subspace is called the $\boldsymbol\Sigma$-envelope of $span(\boldsymbol\beta)$ in full and the envelope for brevity.  $\textbf{P}_{\boldsymbol\Gamma_{1}} \textbf{Y}$ and $\textbf{Q}_{\boldsymbol\Gamma_{0}} \textbf{Y}$ are referred as material and immaterial parts of $\textbf{Y}$, respectively, where $u \leq r$, is referred as the  dimension of the envelope subspace. 

Following the envelope idea, model (\ref{en111}) can be rewritten as
\begin{equation}
\label{spn1}
\textbf{Y}=\boldsymbol\alpha+\boldsymbol\Gamma_{1}\boldsymbol\eta\textbf{X}+\boldsymbol\epsilon,
\end{equation}
where $\boldsymbol\beta=\boldsymbol\Gamma_{1}\boldsymbol\eta,~\boldsymbol\eta\in \mathbbm{R}^{u\times p}$, and $\boldsymbol\Sigma= \boldsymbol\Sigma_{0} + \boldsymbol\Sigma_{1}$ such that $\boldsymbol\Sigma_{0} = \textbf{Q}_{\boldsymbol\Gamma_{1}}\boldsymbol\Sigma\textbf{Q}_{\boldsymbol\Gamma_{1}}^{T}$ being the variance of the immaterial part of response and $\boldsymbol\Sigma_{1} = \textbf{P}_{\boldsymbol\Gamma_{1}}\boldsymbol\Sigma\textbf{P}_{\boldsymbol\Gamma_{1}}^{T}$ being the variance of the material part of response. \cite{cook2010envelope} showed that $\boldsymbol\Sigma = \boldsymbol\Gamma_{1}\boldsymbol\Omega_{1}\boldsymbol\Gamma_{1}^T+\boldsymbol\Gamma_{0}\boldsymbol\Omega_{0}\boldsymbol\Gamma_{0}^T$
where $\boldsymbol\Omega_{1} = var (\boldsymbol\Gamma_{1}^{T}\textbf{Y}) \in \mathbbm{R}^{u\times u}$ and $\boldsymbol\Omega_0 = var (\boldsymbol\Gamma_{0}^{T}\textbf{Y}) \in \mathbbm{R}^{(r-u)\times (r-u)}$ are unknown positive definite matrices with $0 < u \leq r$. Here, one only needs to estimate $r+pu+r(r+1)/2$ parameters. The difference in the number of parameters between the envelope and classical multivariate regression is $p(r-u)$. More details can be found in \cite{cook2010envelope} and the references therein.


\section{New Spatial Envelope} 
In this section, we detail the spatial envelope method. We start with a review of spatial multivariate model, then derive the likelihood function of spatial envelope model, and show the computational steps for the parameter estimation. Let $Y(s_{i}) = (y_1 (s_{i}),\ldots, y_r(s_{i}))^{T}$ be an $r$-variate stochastic spatial response vector along with $p$ regressors $X(s_{i}) =  (x_1(s_{i}),\ldots, x_p(s_{i}))^{T}$ observed at locations $s=\{s_{1},s_{2},\ldots,s_{n};~s_{i} \in \mathbbm{R}^{2}; i =1, 2, \ldots, n\}$. The multivariate spatial regression model can be written as: 
\begin{equation}
\label{sp1}
Y(s_{i}) = \boldsymbol\alpha + \boldsymbol\beta X(s_{i})+\boldsymbol\epsilon(s_{i}),
\end{equation}
where $Y(s)$ denotes the $r\times 1$ response vector at location $s_{i}$ for $i=1,\ldots,n$, $X(s)$ is the $p \times 1$ vector  of fixed and nonstochastic covariates. Furthermore,  $\boldsymbol\alpha$ denotes the $r\times 1$ vector of intercept, $\boldsymbol\beta$ is the $r\times p$ matrix of regression coefficients, and $\boldsymbol\epsilon$ is a multivariate spatial process with mean 0. We assume that the data generating process is second order stationary and the covariance of the response vectors $Y(s_{i})$ and $Y(s_{j})$ at two sites $s_i$ and $s_j$ is a function of distance between the two sites. Namely the covariance can be written as:
\begin{equation}
\label{sp2}
Cov(Y(s_{i}),Y(s_{j}))=C_{ij}(\textbf{h}),~~\textbf{h}=||s_{i}-s_{j}||,
\end{equation}
where $||\cdot||$ denotes Euclidean distance. The function $C(\textbf{h})=\left\{C_{ij}(\textbf{h})\right\}$ is the multivariate covariogram, $C_{ij}(\cdot)$ is the direct covariogram for $i= j$ and cross-covariogram for $i \neq j$. By adopting the \textit{proportional correlation model} \citep{chiles2009geostatistics}, the spatial covariance function can be written as 
\begin{equation}
\label{spcor}
C_{ij}(\textbf{h})=\textbf{V}\rho_{ij}(\textbf{h}),
\end{equation}
where $\textbf{V}$ is an $r \times r$ positive definite matrix and $\rho_{ij}(\textbf{h})$ is the spatial correlation between two sits $s_i$ and $s_j$ \citep{wackernagel2013multivariate}. Estimating the correlation function solely from the data without any structural assumptions is difficult and sometimes infeasible. Usually, it is assumed that the form of the correlation function is a known function but with unknown parameters $\boldsymbol\theta$, which control range, smoothness, and other characteristics of the correlation function. Thus instead of $\rho(\textbf{h})$, we use $\rho(\textbf{h},\boldsymbol\theta)$ to represent unknown parameters $\boldsymbol\theta$ in the correlation function.  For simplicity of notation, $\rho(\textbf{h},\boldsymbol\theta)$ is denoted by $\rho(\boldsymbol\theta)$ throughout the rest of the paper. 

The matrix form for model (\ref{sp1}) 
\begin{equation}
\label{sp11}
\textbf{Y}(s) = \boldsymbol\alpha^{T}\otimes \textbf{1}_{n} + \textbf{X}(s)\boldsymbol\beta^{T}+\boldsymbol\epsilon(s),
\end{equation}
where $\textbf{Y}(s) = \begin{pmatrix}
				Y^{T}(s_{1}) \\
				 \vdots \\
				  Y^{T}(s_{n})		\end{pmatrix}$ 
denotes the $n\times r$ response matrix $\textbf{X}(s)$ is the $n \times p$ matrix of covariates. Furthermore, $\otimes$ denotes the Kronecker product and $\textbf{1}_{n}$ is an $n\times 1$ column vector with 1 at each entry. From the envelope idea, $\textbf{V}$ can be written as $\textbf{V}_{0}+\textbf{V}_{1}$ where $\textbf{V}_{0} = \textbf{Q}_{\boldsymbol\Gamma_{1}}\textbf{V}\textbf{Q}_{\boldsymbol\Gamma_{1}}$ denotes the covariance matrix associated with the immaterial part of response and $\textbf{V}_{1}= \textbf{P}_{\boldsymbol\Gamma_{1}}\textbf{V}\textbf{P}_{\boldsymbol\Gamma_{1}}$ denotes the covariance matrix associated with the material part where $\boldsymbol\Gamma_{1}$ is the semi-orthogonal basis of $span(\textbf{V}_{1})$. Hence, the spatial covariance matrix of $C_{ij}(\textbf{h})$ can be written as follows:
\begin{eqnarray}
\label{var1}
C_{ij}(\textbf{h}) &=& \textbf{V}\rho_{ij}(\boldsymbol\theta)\cr
&=&(\textbf{V}_{0}+ \textbf{V}_{1})\rho_{ij}(\boldsymbol\theta)
\end{eqnarray}
Let $0 < u \leq r$ denotes the structural dimension of the envelope, where $u$ can be selected using a modified information criterion such as modified BIC (\cite{li2017parsimonious}), model free dimension selection such as full Grassmanian (FG; \citealp{zhang2017model}) and the 1-D algorithm \citep{cook2016algorithms} , or cross-validation. More details can be found in \citep{zhang2017model, zhang2018functional} and the references therein.

To illustrate the estimation, we use a $vec$ operator on the response matrix. That is, let $\mathbbm{Y} (s) = vec(\textbf{Y}(s))$ be an $nr\times 1$ vector for the vectorized response variable, and $\mathbbm{X}(s) = \textbf{I}_{r}\otimes\textbf{X}(s)$ be an $nr\times pr$ block diagonal matrix having $\textbf{X}_{i}(s)$ as blocks. Thus, the vectorized version of the multivariate spatial linear model can be written as:  
\begin{equation}
\label{model}
\mathbbm{Y}(s) = \boldsymbol\alpha\otimes\textbf{1}_{n}+\mathbbm{X}(s)\boldsymbol\beta^*+ \boldsymbol\epsilon^{*}(s).
\end{equation}
where $\boldsymbol\alpha$ is an $r\times 1$ vector of intercept, $\boldsymbol\beta^*=vec(\boldsymbol\beta^{T})$ shows an $pr\times 1$ vector of regression coefficients, and $\boldsymbol\epsilon^*(s)$ is an $nr\times 1$ vector of spatial errors with mean 0. With the use of proportional covariance model and the vectorization of the response matrix, the $nr\times nr$ covariance matrix of the response variables $\boldsymbol\Sigma_{\mathbbm{Y}}$, can be written as $\textbf{V}\otimes \boldsymbol\rho(\boldsymbol\theta)$.

The likelihood function of model (\ref{model}) is:
\begin{equation}
\label{lik1}
\begin{aligned}
&L(\boldsymbol\alpha,\boldsymbol\beta^{*},\textbf{V},\boldsymbol\theta)=\left[\det(\textbf{V}\otimes \boldsymbol\rho(\boldsymbol\theta))\right]^{-\frac{1}{2}}\\
&\times exp\left\{-\frac{1}{2}(\mathbbm{Y}(s)-\boldsymbol\alpha\otimes\textbf{1}_{n}- \mathbbm{X}(s)\boldsymbol\beta^{*})^{T}(\textbf{V}\otimes \boldsymbol\rho(\boldsymbol\theta))^{-1}(\mathbbm{Y}(s)-\boldsymbol\alpha\otimes\textbf{1}_{n}- \mathbbm{X}(s)\boldsymbol\beta^{*})\right\},
\end{aligned}
\end{equation}
where $\det(\cdot)$ denotes the determinant of the matrix. Suppose the response vector can be decomposed into the material and immaterial part, $\mathbbm{Y}_{1} = (\textbf{I}_{r}\otimes \textbf{P}_{\boldsymbol\Gamma_{1}})\mathbbm{Y}(s) $ and $\mathbbm{Y}_{0}= (\textbf{I}_{r}\otimes \textbf{Q}_{\boldsymbol\Gamma_{1}})\mathbbm{Y}(s)$, respectively. 
From (\ref{var1}), the covariance matrix of $\mathbbm{Y}(s)$ can be written as follows:
\begin{equation}
\label{var11}
\begin{aligned}
\boldsymbol\Sigma_{\mathbbm{Y}}&=\textbf{V}\otimes\boldsymbol\rho(\boldsymbol\theta)\cr
&=\textbf{V}_{0}\otimes\boldsymbol\rho(\boldsymbol\theta)+\textbf{V}_{1}\otimes\boldsymbol\rho(\boldsymbol\theta).
\end{aligned}
\end{equation}
Combining (\ref{lik1}) and (\ref{var11}), we have 
\begin{equation}
\label{lik222}
L^{u}(\boldsymbol\alpha, \boldsymbol\beta^{*},\textbf{V}_{0},\textbf{V}_{1},\boldsymbol\theta)=L_{1}^{u}(\boldsymbol\alpha, \boldsymbol\beta^{*},\textbf{V}_{1},\boldsymbol\theta)\times L_{2}^{u}(\boldsymbol\alpha, \textbf{V}_{0},\boldsymbol\theta),
\end{equation}
with 
\begin{equation}
\label{lik4}
\begin{aligned}
&L_{1}^{u}(\boldsymbol\alpha, \boldsymbol\beta^{*},\textbf{V}_{1},\boldsymbol\theta)= [det_{0}(\textbf{V}_{1})]^{-\frac{n}{2}}[\det(\boldsymbol\rho(\boldsymbol\theta))]^{-\frac{r}{2}} \\
&\times exp\left\{-\frac{1}{2}\left(\mathbbm{Y}(s)-\boldsymbol\alpha\otimes\textbf{1}_{n}- \mathbbm{X}(s)\boldsymbol\beta^{*}\right)^{T} \left(\textbf{V}_{1}^{\dagger}\otimes\boldsymbol\rho^{-1}(\boldsymbol\theta)\right) \left(\mathbbm{Y}(s)-\boldsymbol\alpha\otimes\textbf{1}_{n}- \mathbbm{X}(s) \boldsymbol\beta^{*}\right)\right\},\\
&L_{2}^{u}(\boldsymbol\alpha, \textbf{V}_{0},\boldsymbol\theta) = [det_{0}(\textbf{V}_{0})]^{-\frac{n}{2}}[\det(\boldsymbol\rho(\boldsymbol\theta))]^{-\frac{r}{2}}\\
&\times exp\left\{-\frac{1}{2}(\mathbbm{Y}(s)-\boldsymbol\alpha\otimes\textbf{1}_{n})^{T} \left(\textbf{V}_{0}^{\dagger}\otimes\boldsymbol\rho^{-1}(\boldsymbol\theta)\right) (\mathbbm{Y}(s)-\boldsymbol\alpha\otimes\textbf{1}_{n})\right\},
\end{aligned}
\end{equation}
where $\dagger$ denotes the Moore-Penrose inverse and $det_{0}(\textbf{A})$ denotes the product of non-zero eigenvalues of a non-zero symmetric matrix \textbf{A}. The likelihood in equation (\ref{lik1}) can be factorized as equation (\ref{lik222}) from $span(\boldsymbol\beta) \subseteq span(\textbf{V}_{1})$, and $(\textbf{V}_{0}^{\dagger}\otimes\boldsymbol\rho^{-1}(\boldsymbol\theta))\mathbbm{X}\boldsymbol\beta^{*}= \textbf{0}$. This factorization is detailed in the Appendix, section 9.1. 

The objective is to maximize the likelihood in (\ref{lik222}) over $\boldsymbol\beta^{*},\textbf{V}_{0},\textbf{V}_{1}$, and $\boldsymbol\theta$ subject to the constraints:
\begin{equation}
\begin{aligned}
\label{con}
\begin{split}
&span(\boldsymbol\beta) \subseteq span(\textbf{V}_{1}),\\
&\textbf{V}_{0}\textbf{V}_{1} = 0.
\end{split}
\end{aligned}
\end{equation}
Thus, the multivariate spatial model in (\ref{model}) can be written as 
\begin{equation}
\begin{aligned}
\label{model2}
\mathbbm{Y}(s) &= \boldsymbol\alpha\otimes\textbf{1}_{n}+\mathbbm{X}(s)vec(\boldsymbol\eta^{T}\boldsymbol\Gamma_{1}^{T})+ \boldsymbol\epsilon^{*}(s),\\
\boldsymbol\Sigma &= \left(\boldsymbol\Gamma_{1}\boldsymbol\Omega_{1}\boldsymbol\Gamma_{1}^{T}+\boldsymbol\Gamma_{0}\boldsymbol\Omega_{0}\boldsymbol\Gamma_{0}^{T}\right) \otimes \boldsymbol\rho(\boldsymbol\theta),
\end{aligned}
\end{equation}
where $\boldsymbol\Gamma_{1}$ denotes the semi-orthogonal basis for $span(\textbf{V}_1)$, $\boldsymbol\Gamma_{0}$ denotes the semi-orthogonal basis for the orthogonal complement space of $span(\textbf{V}_1)$, $\boldsymbol\Omega_{1}$ denotes the covariance of the material part of response, $\boldsymbol\Omega_{2}$ denotes the covariance of the immaterial part of response, and $\boldsymbol\eta \in \mathbbm{R}^{u\times r} $ is chosen such that $\boldsymbol\beta^{*} = vec(\boldsymbol\eta^{T}\boldsymbol\Gamma_{1}^{T})$. 

As mentioned by \cite{cook2010envelope}, the gradient-based algorithms for Grassmann optimization \citep{edelman1998geometry} require a coordinate version of the objective function which must have continuous directional derivatives. The optimization depends on minimizing the logarithm of $\textbf{D}$ over the Grassmann manifold $\mathbbm{G}^{r\times u}$, where 
\begin{equation*}
\textbf{D} =  \det(\textbf{P}_{\textbf{V}_{1}}\hat{\boldsymbol\Sigma}_{\textbf{res}}\textbf{P}_{\textbf{V}_{1}}+ \textbf{Q}_{\textbf{V}_{1}}\hat{\boldsymbol\Sigma}_{\mathbbm{Y}}\textbf{Q}_{\textbf{V}_{1}}),
\end{equation*}

\noindent and $\textbf{D}$ is the partially maximized likelihood function. The derivation of $\textbf{D}$ is detailed in the Appendix, section 9.2. Let $\hat{\boldsymbol\Gamma}_{1}$ be the semi-orthogonal basis for $span(\textbf{V}_1)$ and $\hat{\boldsymbol\Gamma}_{0}$ be the semi-orthogonal basis for $span(\textbf{V}_0)$. Then $\hat{\boldsymbol\eta} = \hat{\boldsymbol\Gamma}_{1}^{T}\hat{\boldsymbol\beta}$, $\hat{\boldsymbol\Omega}_{1}= \hat{\boldsymbol\Gamma}_{1}^{T} \hat{\boldsymbol\Sigma}_{\textbf{res}}\hat{\boldsymbol\Gamma}_{1}$ and $\hat{\boldsymbol\Omega}_{0}= \hat{\boldsymbol\Gamma}_{0}^{T} \hat{\boldsymbol\Sigma}_{\mathbbm{Y}}\hat{\boldsymbol\Gamma}_{0}$, where $\hat{\boldsymbol\Sigma}_{\mathbbm{Y}}$ and $\hat{\boldsymbol\Sigma}_{\textbf{res}}$ are the marginal covariance matrix of $\mathbbm{Y}$ and the residual covariance matrix, respectively. Let $\log \det(\cdot)$ denote the composite function $\log \circ \det(\cdot)$. Then, the coordinate form of the $\log \textbf{D}$ 

\begin{scriptsize}
\begin{equation}
\label{logd1}
\log \textbf{D} =\log \det\left(\boldsymbol\Gamma_{1}^{T}\left( \textbf{H}^{T} \hat{\boldsymbol\rho}^{-1}(\boldsymbol\theta)\textbf{H} - \textbf{H}^{T} \hat{\boldsymbol\rho}^{-1}(\boldsymbol\theta)\textbf{G} \left(\textbf{G}^{T} \hat{\boldsymbol\rho}^{-1}(\boldsymbol\theta)\textbf{G} \right)^{-1}\textbf{G}^{T} \hat{\boldsymbol\rho}^{-1}(\boldsymbol\theta)\textbf{H}\right)\boldsymbol\Gamma_{1}+ \boldsymbol\Gamma_{0}^{T}(\textbf{H}^{T} \hat{\boldsymbol\rho}^{-1}(\boldsymbol\theta)\textbf{H})\boldsymbol\Gamma_{0} \right)
\end{equation}
\end{scriptsize}

\noindent where $\textbf{H}=\textbf{Y}-\bar{\textbf{Y}}\otimes\textbf{1}_{n}$, and $\textbf{G} = \textbf{X}-\bar{\textbf{X}}\otimes \textbf{1}_{n}$. 

 In order to obtain the parameters of spatial envelope model, the objective function (\ref{logd1}) can be minimized by the gradient based Grassmann optimization. To do this, first obtain an initial value for $\hat{\boldsymbol\Sigma}_{\textbf{Y}}^{0}$, $\hat{\boldsymbol\Sigma}_{\textbf{res}}^{0}$, and $\hat{\boldsymbol\beta}_{MLE}$, the marginal covariance matrix of $\mathbbm{Y}$, the residual covariance matrix, and the maximum likelihood estimate for $\boldsymbol\beta$ from the fit of the full model (\ref{model}). Set $\boldsymbol\Theta^{1} = \boldsymbol\Theta^{0}$ where $\boldsymbol\Theta = \{\boldsymbol\theta, \textbf{V}_{0}, \textbf{V}_{1}\}$ and $\textbf{V}_{0}$ and $\textbf{V}_{1}$ can be obtained using traditional envelope model and $\boldsymbol\theta$ can be obtained using linear coregionalization model. Then, we estimate $\textbf{P}_{\textbf{V}_{1}^{m}}$ by minimizing the objective function (\ref{logd1}) over the Grassmann manifold $\mathbbm{G}^{(r\times u)}$, and estimate $\textbf{P}_{\textbf{V}_{0}^{m}}$ by $\hat{\textbf{P}}_{\textbf{V}_{0}^{m}} = \textbf{I} - \hat{\textbf{P}}_{\textbf{V}_{1}^{m}} $. In order to update the covariance function of material and immaterial parts of the spatial envelope, fix $\boldsymbol\theta^{m}$ and estimate $\textbf{V}_{0}^{m}$ and $\textbf{V}_{1}^{m}$ by $\hat{\textbf{V}_{0}^{m}} =\hat{\textbf{P}}_{\textbf{V}_{0}^{m}}\hat{\boldsymbol\Sigma}_{\textbf{Y}}^{m} \hat{\textbf{P}}_{\textbf{V}_{0}^{m}} $ and $\hat{\textbf{V}_{1}^{m}} = \hat{\textbf{P}}_{\textbf{V}_{1}^{m}}\hat{\boldsymbol\Sigma}_{\textbf{res}}^{m} \hat{\textbf{P}}_{\textbf{V}_{1}^{m}}$. Then, fix $\textbf{V}_{0}^{m}$ and $\textbf{V}_{1}^{m}$ and maximize $L^{(u)}(\boldsymbol\alpha, \boldsymbol\beta,\textbf{V}_{0}^{m},\textbf{V}_{1}^{m},\boldsymbol\theta^{m})$ over $\boldsymbol\theta$ by solving the following minimization problem using numerical algorithm such as Newton-Raphson method:

\begin{scriptsize}
\begin{equation}
\begin{aligned}
\hat{\boldsymbol\theta^{m}}&=\argmaxF_{\boldsymbol\theta} \{r \det(\boldsymbol\rho(\boldsymbol\theta))+\\
&\frac{1}{2}tr\left(\left( \textbf{Q}_{\left(\boldsymbol\rho^{-\frac{1}{2}}(\boldsymbol\theta)\textbf{G}\right)}\boldsymbol\rho^{-\frac{1}{2}}(\boldsymbol\theta)\textbf{H}\right) \textbf{V}_{1}^{{m}^\dagger}\left(\textbf{Q}_{\left(\boldsymbol\rho^{-\frac{1}{2}}(\boldsymbol\theta)\textbf{G}\right)}\boldsymbol\rho(\boldsymbol\theta)^{-\frac{1}{2}}\textbf{H}\right)^{T} + \boldsymbol\rho^{-\frac{1}{2}}(\boldsymbol\theta)\textbf{H} \textbf{V}_{0}^{{m}^\dagger} \textbf{H}^{T} \boldsymbol\rho^{-\frac{1}{2}}(\boldsymbol\theta)\right) \}.
\end{aligned}
\end{equation}
\end{scriptsize}

\noindent Now, update $\hat{\boldsymbol\Sigma}_{\textbf{Y}}^{m}$ and $\hat{\boldsymbol\Sigma}_{\textbf{res}}^{m}$ using the new estimate for $\textbf{V}_{0},\textbf{V}_{1}$, and $\boldsymbol\theta$. Then, check the convergence. If $||\boldsymbol\Theta^{m+1}-\boldsymbol\Theta^{m}||< \delta$ where  $\delta$ is a pre-specified tolerance level, then stop the iteration, output the final spatial envelope estimators and estimate $\boldsymbol\beta$ by $\hat{\boldsymbol\beta}=\hat{\textbf{P}}_{\textbf{V}_{1}}\hat{\boldsymbol\beta}_{MLE}$; otherwise, set $m := m + 1$ and redo the procedure. Finally, estimate the intercept by $\hat{\boldsymbol\alpha}= \bar{\textbf{Y}}-\bar{\textbf{X}}\hat{\boldsymbol\beta}^{T}$. When the problem reduces to a standard envelope estimation problem, the fast algorithm for the envelope such as \cite{cook2016note} can be applied.


\section{Theoretical Properties}
In what follows, we study the asymptotic properties of the spatial envelope parameter estimates. The regression coefficients can be written as $\boldsymbol\beta = \boldsymbol\Gamma_{1}\boldsymbol\eta$. Furthermore, $\textbf{V}_{0}=\boldsymbol\Gamma_{0}\boldsymbol\Omega_{0}\boldsymbol\Gamma_{0}^{T}$ and $\textbf{V}_{1}=\boldsymbol\Gamma_{1}\boldsymbol\Omega_{1}\boldsymbol\Gamma_{1}^{T}$ are the covariance of the immaterial part and material part to the regression, respectively. Therefore, aside from the intercept, the parameters of spatial envelope model in equation (\ref{model}) can be combined into the vector as follows: 
\begin{equation}
\label{asym1}
\boldsymbol\phi = \begin{bmatrix}
    vec(\boldsymbol\eta)\\
    vec(\boldsymbol\Gamma_{1})\\
    vech(\boldsymbol\Omega_{1})\\
    vech(\boldsymbol\Omega_{0})\\
\end{bmatrix}
\equiv \begin{bmatrix}
    \phi_{1}\\
    \phi_{2} \\
    \phi_{3} \\
    \phi_{4} \\
\end{bmatrix}
\end{equation}
where the $vec(\cdot)$ denotes the vector operator and $vech(\cdot)$ denotes vector half operator. For background on these operators, see \cite{seber2008matrix}. Here we focus on the following parameters under the spatial envelope model:
\begin{equation}
\label{asym2}
\psi(\boldsymbol\phi) = \begin{bmatrix}
   vec(\boldsymbol\beta^{*})\\
    vech(\textbf{V})\\
\end{bmatrix}
= \begin{bmatrix}
    vec(\boldsymbol\eta^{T}\boldsymbol\Gamma_{1}^{T})\\
    vech\left((\boldsymbol\Gamma_{1}\boldsymbol\Omega_{1}\boldsymbol\Gamma_{1}^{T}+\boldsymbol\Gamma_{0}\boldsymbol\Omega_{0}\boldsymbol\Gamma_{0}^{T})\right)\\
\end{bmatrix}
\equiv \begin{bmatrix}
    \psi_{1}(\boldsymbol\phi)\\
    \psi_{2}(\boldsymbol\phi) \\
\end{bmatrix}
\end{equation}
Let
\begin{equation}
\label{asym3}
\Psi = \begin{bmatrix}
\frac{\partial \psi_{1}}{\partial  \phi_{1}^{T}}& \ldots &\frac{\partial \psi_{1}}{\partial  \phi_{4}^{T}}\\
\frac{\partial \psi_{2}}{\partial  \phi_{1}^{T}}& \ldots &\frac{\partial \psi_{2}}{\partial  \phi_{4}^{T}}\\
\end{bmatrix}
\end{equation}
denote the gradient matrix. Using this gradient matrix and following \cite{cook2010envelope}, we present the following asymptotic properties of proposed estimators.

\textbf{Lemma 1:} Suppose $\bar{\textbf{X}} = 0$, the Fisher information, $\textbf{J}$, for $\psi(\boldsymbol\phi)$ in the model (\ref{model}) is as follows:
\begin{equation}
\label{asym4}
\begin{aligned}
\textbf{J}&= \begin{bmatrix}
\frac{1}{n}\mathbbm{X}^{T}\left( \textbf{V}^{-1}\otimes\boldsymbol\rho^{-1}(\boldsymbol\theta)\right)\mathbbm{X} & \textbf{0}\\
\textbf{0} &  \frac{1}{2}\textbf{E}^T_{r}\left(\textbf{V}^{-1}\otimes\textbf{V}^{-1}\right)\textbf{E}_{r}\\
\end{bmatrix}
\\
&=  
\begin{bmatrix}
\textbf{V}^{-1} \otimes \left(\frac{\textbf{X}^{T}\boldsymbol\rho^{-1}(\boldsymbol\theta)\textbf{X}}{n}\right) & \textbf{0}\\
\textbf{0} &  \frac{1}{2}\textbf{E}^T_{r}\left(\textbf{V}^{-1}\otimes\textbf{V}^{-1}\right)\textbf{E}_{r}\\
\end{bmatrix}.
\end{aligned}
\end{equation}
where $\textbf{E}_{r}\in R^{r^{2}\times r(r+1)/2}$ is an expansion matrix such that for a matrix $\textbf{A}$, $vec(\textbf{A}) = \textbf{E}_{r}vech(\textbf{A})$, and $diag(\textbf{A})$ is the matrix with the diagonal elements of $\textbf{A}$.  The derivation of \textbf{J} is provided in the Appendix, section 9.3. 

\textbf{Theorem 1:} Suppose $\bar{\textbf{X}} = 0$ and $\textbf{J}$ is the Fisher information defined in lemma 1. Let $\boldsymbol\Lambda = \textbf{J}^{-1}$ be the asymptotic variance of the MLE under the full model. Then
\begin{equation}
\label{asym5}
\sqrt{n}(\hat{\boldsymbol\phi}-\boldsymbol\phi)\rightarrow N(\textbf{0},\boldsymbol\Lambda_{0})
\end{equation}
where $\boldsymbol\Lambda_{0} = \Psi(\Psi^{T}\boldsymbol\Lambda\Psi)^{\dagger}\Psi$. Furthermore, $\boldsymbol\Lambda^{-\frac{1}{2}}(\boldsymbol\Lambda-\boldsymbol\Lambda_{0})\boldsymbol\Lambda^{-\frac{1}{2}} \geq 0$, which means the asymptotic variance of the parameter estimation under the spatial envelope model is smaller than their estimate under MLE. Proof of this theorem can be found in the Appendix, section 9.4.

\textbf{Corollary 1:} The asymptotic variance (avar) of $\sqrt{n}\boldsymbol\beta^*$ can be written as
\begin{equation}
\label{asymp7}
avar(\sqrt{n}\boldsymbol\beta^*) = K_{rp}\left\{\left(\frac{\textbf{X}^{T}\boldsymbol\rho(\boldsymbol\theta)^{-1}\textbf{X}}{n}\right)^{-1}\otimes \boldsymbol\Gamma_{1}\boldsymbol\Omega_{1}\boldsymbol\Gamma_{1}^{T} + (\boldsymbol\eta^{T}\otimes \boldsymbol\Gamma_{0})(\Psi_{2}^{T}\textbf{J}\Psi_{2})^{\dagger}(\boldsymbol\eta\otimes \boldsymbol\Gamma_{0}^{T})\right\}K_{rp}^T
\end{equation}
where $\Psi_{2} = \left(\frac{\partial \psi_{1}}{\partial  \phi_{2}^{T}}, \frac{\partial \psi_{2}}{\partial  \phi_{2}^{T}}\right)^{T}$ and $K_{rp} \in \mathbbm{R}^{rp\times rp}$ is the unique matrix such that for a matrix $\textbf{A}$, $vec(\textbf{A}^{T}) =K_{rp}vec(\textbf{A})$ i.e. $K_{rp}$ transforms the $vec$ of a matrix into the $vec$ of its transpose. Proof of this theorem can be found in the Appendix, section 9.5.

To gain further insight into the structure of the spatial envelope, we present the simply version of the asymptotic variance of the $\boldsymbol\beta^*$ for the cases that we have one covariate, $\boldsymbol\Omega_{1} = \sigma_{1}^{2}\textbf{I}_{u}$, and $\boldsymbol\Omega_{0} = \sigma_{0}^{2}\textbf{I}_{r-u}$. Then, the asymptotic variance of the $\boldsymbol\beta^*$ can be shown to be 
\begin{eqnarray}
\label{asymbet}
avar(\sqrt{n}\boldsymbol\beta^{*}) = \frac{n\sigma_{1}^{2}}{\textbf{X}^{T}\boldsymbol\rho^{-1}(\boldsymbol\theta)\textbf{X}}\boldsymbol\Gamma_{1}\boldsymbol\Gamma_{1}^{T} +
\frac{n\sigma_{0}^{2}\sigma_{1}^{2}||\boldsymbol\beta||^{2}}{\textbf{X}^{T}\boldsymbol\rho^{-1}(\boldsymbol\theta)\textbf{X}\sigma_{1}^{2}
||\boldsymbol\beta||^{2}+n(\sigma_{0}^{2}-\sigma_{1}^{2})^{2}}\boldsymbol\Gamma_{0}\boldsymbol\Gamma_{0}^{T}. ~~
\end{eqnarray}
For this simplify version, it can be shown that 
\begin{eqnarray}
\label{comp}
\frac{\textbf{V}_{SPEN}^{-\frac{1}{2}}\textbf{V}_{EN}\textbf{V}_{SPEN}^{-\frac{1}{2}}}{\frac{\textbf{X}^{T}\boldsymbol\rho^{-1}(\boldsymbol\theta)\textbf{X}}{n\sigma_{\textbf{X}}^{2}}} =
\textbf{I}_r +  \left( \frac{(\sigma_{0}^{2}-\sigma_{1}^{2})^{2}\left(\frac{n\sigma_{\textbf{X}}^{2}}{\textbf{X}^{T}\boldsymbol\rho^{-1}(\boldsymbol\theta)\textbf{X}}-1\right)}
{(\sigma_{0}^{2}-\sigma_{1}^{2})^{2}+\sigma_{1}^{2}\sigma_{\textbf{X}}^{2}||\boldsymbol\beta||^{2}}
\right)\boldsymbol\Gamma_{0}\boldsymbol\Gamma_{0}^{T},~
\end{eqnarray}
where $\textbf{V}_{SPEN}$ shows the asymptotic variance of the spatial envelope model, $\textbf{V}_{EN}$ shows the asymptotic variance of the envelope model, and $\sigma_{\textbf{X}}^{2}$ denotes the variance of the $\textbf{X}$ which is an $n\times1$ vector. Proof of equation (\ref{comp}) can be found in the Appendix, section 9.6.  
This results indicates that when the spatial correlation does not exists, i.e. $\boldsymbol\rho(\boldsymbol\theta)=\textbf{I}$, the asymptotic variance for both model would be equal. On the other hand, for the cases that spatial correlation exists, drawing an analytical conclusion for comparing the asymptotic variance of the two models is very difficult. In this case, the variance of the two models can be compared numerically. oth model would be equal.


\setcounter{equation}{0} 
\section{Prediction}
Prediction at an unsampled location is often a major objective of a spatial analysis. Let $\mathbbm{Y}_{new}$ be the $vec(\textbf{Y}_{new})$ of the new multivariate response and $\mathbbm{X}_{new}$ be the predictor vector at an unsampled location. The model then can be written as:
\begin{equation}
 \begin{pmatrix} \mathbbm{Y}_{new} \\  \mathbbm{Y} \end{pmatrix} =  \begin{pmatrix}\boldsymbol\alpha\otimes\textbf{1}_{n_{new}}+\mathbbm{X}_{new}\boldsymbol\beta^{*} \\ \boldsymbol\alpha\otimes\textbf{1}_{n}+\mathbbm{X}\boldsymbol\beta^{*} \end{pmatrix} +  \begin{pmatrix} \boldsymbol\epsilon_{new} \\ \boldsymbol\epsilon \end{pmatrix}\sim N\left( \boldsymbol\alpha\otimes\textbf{1}_{N}+\begin{pmatrix}\mathbbm{X}_{new} \\ \mathbbm{X} \end{pmatrix} \boldsymbol\beta^{*},\boldsymbol\Sigma\right).
\end{equation}
where $N=n+n_{new}$ and $\boldsymbol\Sigma$ is as follows
\begin{equation}
 \boldsymbol\Sigma = \begin{pmatrix} \boldsymbol\Sigma_{11} & \boldsymbol\Sigma_{12}\\  \boldsymbol\Sigma_{21} & \boldsymbol\Sigma_{22} \end{pmatrix} =   \begin{pmatrix} (\textbf{V}_{0}+\textbf{V}_{1})\otimes\boldsymbol\rho_{new,new}(\boldsymbol\theta) & (\textbf{V}_{0}+\textbf{V}_{1})\otimes\boldsymbol\rho_{new,\textbf{Y}}(\boldsymbol\theta)\\  (\textbf{V}_{0}+\textbf{V}_{1})\otimes\boldsymbol\rho_{\textbf{Y},new}(\boldsymbol\theta) & (\textbf{V}_{0}+\textbf{V}_{1})\otimes\boldsymbol\rho_{\textbf{Y},\textbf{Y}}(\boldsymbol\theta) \end{pmatrix} .
\end{equation}
The conditional distribution $\textbf{Y}_{new}|\textbf{Y}$ is 
\begin{equation}
\label{pred}
\textbf{Y}_{new}|\textbf{Y}, \boldsymbol\alpha,\boldsymbol\eta, \textbf{V}_{0}, \textbf{V}_{1}, \boldsymbol\theta\sim  N\left(\boldsymbol\mu_{1}+\boldsymbol\Sigma_{12}\boldsymbol\Sigma_{22}^{-1}(\textbf{Y}-\boldsymbol\mu_{2}),\boldsymbol\Sigma_{11}-\boldsymbol\Sigma_{12}\boldsymbol\Sigma_{22}^{-1}\boldsymbol\Sigma_{21}\right),
\end{equation}
where $\boldsymbol\mu_{1}=\boldsymbol\alpha\otimes\textbf{1}_{n_{new}}+\mathbbm{X}_{new}\boldsymbol\beta^{*}$ and $\boldsymbol\mu_{2}=\boldsymbol\alpha\otimes\textbf{1}_{n}+\mathbbm{X}\boldsymbol\beta^{*}$. Using the method described in section 3, one can estimate the parameters of the model and then from the conditional distribution (\ref{pred}) the $E(\textbf{Y}_{new}|\textbf{Y})$ can be estimated.


\section{Simulation }
In this section, we carry out a simulation study to evaluate the finite sample performance of the proposed spatial envelope model and to compare it with the traditional multivariate linear regression (MLR), linear coregionalization model (LCM; \citealp{zhang2007maximum}), and envelope \citep{cook2010envelope}. 

The data $\left\{(\textbf{X}_1,\textbf{Y}_1),\ldots, (\textbf{X}_n,\textbf{Y}_n)\right\}$ are generated from the model 
\begin{equation}
\label{simmod}
\textbf{Y}=\textbf{X}\boldsymbol\beta + \boldsymbol\epsilon, \\
\end{equation}
where $\textbf{Y}_i \in \mathbbm{R}^{5}$, $\textbf{X}_i \in \mathbbm{R}^{6}$, and the structural dimension $u=2$. The matrix $(\boldsymbol\Gamma_{1};\boldsymbol\Gamma_{0})$ is obtained by orthogonalizing an $5\times5$ matrix generated from uniform $(0, 1)$ variables. The elements of $\boldsymbol\eta$ follow standard normal distribution, and $\boldsymbol\beta=\boldsymbol\Gamma_{1}\boldsymbol\eta$. We generate $\boldsymbol\Sigma_{Y} = \left( \boldsymbol\Gamma_{1}\boldsymbol\Omega_{1}\boldsymbol\Gamma_{1}^{T} + 5\boldsymbol\Gamma_{0}\boldsymbol\Omega_{0}\boldsymbol\Gamma_{0}^{T} \right) \otimes \boldsymbol\rho(\boldsymbol\theta)$ where $\boldsymbol\Omega_{1}=[\{ (-0.9)^{|i-j|}\}$ and $\boldsymbol\Omega_{0}=\{ (-0.5)^{|i-j|}\}$. For the spatial correlation function $\boldsymbol\rho(\boldsymbol\theta)$, we use the following  Matern covariance function:
\[
\boldsymbol\rho(h;\boldsymbol\theta)=\frac{\sigma_m^2}{2^{\theta_{2}-1}\Gamma(\theta_{2})}\left(\frac{||h||}{\theta_{1}}\right)^{\theta_{1}}\kappa_{\theta_{2}}\left(\frac{||h||}{\theta_{1}}\right), 
\]
where $\boldsymbol\theta=(\theta_{1},\theta_{2})$, $\theta_{1}>0$ is the range parameter, $\theta_{2}$ is the smoothness parameter, $\Gamma(\cdot)$ is the Gamma function, and $\kappa_{\theta_{2}}$ is the modified Bessel function of the second kind of order $\theta_{2}$ \citep{abramowitz1964handbook}. Three error distributions of $\boldsymbol\epsilon$ are investigated. We assume $\boldsymbol\epsilon$ follows a normal distribution with mean 0 and covariance $\boldsymbol\Sigma$. For first error scenario, $\boldsymbol\Sigma = \left(\boldsymbol\Gamma_{1}\boldsymbol\Omega_{1}\boldsymbol\Gamma_{1}^{T} +5\boldsymbol\Gamma_{0}\boldsymbol\Omega_{0}\boldsymbol\Gamma_{0}^{T}\right)$. This density serves as a benchmark where the errors are independent from each other. For the second scenario, let $\boldsymbol\epsilon$ follows a Matern covariance function with $\sigma_m=3$, $\theta_1=1$, and $\theta_2=0.5$; This case represents a spatial correlation in the data with a short range of dependency. This case is an example of weak spatial correlation. Finally, let $\boldsymbol\epsilon$ follows a Matern covariance function with $\sigma_m=3$, $\theta_1=5$, and $\theta_2=0.5$; This case represents a spatial correlation in the data with a long range of dependency. This case is an example of strong spatial correlation. 

Sample size is 100, 225, and 400. There are two different ways to generate these samples. One is based on $10\times 10$, $15\times 15$ and $20\times 20$ evenly spaced grids on $[0,1]^{2}$, respectively. Another way is to randomly choose 100, 225, and 400 locations from a $101\times 101$ grid on $[0,1]^{2}$. We use both sampling procedures to check whether the spatial distribution of the observations has any impact on the proposed estimation. All results reported here are based on 200 replications from the simulation model in each scenario. In order to compare the different estimators, we use \textit{Leave One Out Cross-Validation} (LOCV) method, which provides a convenient approximation for the prediction error under squared-error loss
\begin{eqnarray}
MSPE=\frac{\sum_{i=1}^{n}(\hat{\textbf{Y}}^{(-i)}(s_{i})-\textbf{Y}(s_{i,obs}))(\hat{\textbf{Y}}^{(-i)}(s_{i})-\textbf{Y}(s_{i,obs}))^T}{n},
\end{eqnarray}
where $\textbf{Y}(s_{i,obs})$ is the observe value for response in location $s_i$ and $\hat{\textbf{Y}}^{(-i)}(s_{i})$ is the predicted values of $\textbf{Y}(s_{i})$ computed with the $i$th row of the data removed.  The Matlab package Envlp was used for all our simulation studies. Tables 1 and 2 summarize the results of these simulations. These tables provide the LOCV for different methods and different error distributions.

\begin{table}[htp]
{\footnotesize
\begin{center}
\label{tab11} \caption{ Prediction accuracy comparison based on the mean (standard deviation) of leave one out cross-validation (LOCV) for all 200 data sets from equally spaced samples. Smaller LOCV shows better performance.} 
\begin{tabular}{|l|l|l|l|l|l| }
 \hline 
 $\epsilon$ & n & MLR &LCM&  Envelope & Spatial Envelope   \\
 \hline 
1&100	&19.02 (1.537)	&20.01 (1.754)	&13.71 (1.547)	& 14.28 (1.644)\\
&225	&18.49 (1.153)	&19.75 (1.659)	&11.49 (1.124)	& 12.51 (1.234)\\
&400	&18.27 (0.828)	&19.02 (1.002)	&10.37 (0.812)	& 10.87 (0.989)\\
\hline
2&100	&102.79 (35.570)	&22.54 (3.246)	&91.98 (36.379)	&20.21 (1.988)\\
&225	&101.57 (32.495)	&20.46 (2.897)	&89.24 (33.083)	&18.34 (1.450)\\
&400	&99.98 (32.185)	&18.89 (2.051)	&88.95 (31.855)	&17.68 (1.056)\\
\hline
3&100	&117.79 (48.834)	&24.19 (4.125)	&119.08 (47.852)	&21.36 (2.353)\\
&225	&103.22 (39.065)	& 21.78 (3.278)	&104.73 (39.023)	&20.76 (2.012)\\
&400	&99.08 (37.718)	&19.45 (3.001)	&100.39 (36.896)	&18.10 (1.651)\\
\hline															
\end{tabular}
\end{center}}
\end{table}

\begin{table}[htp]
{\footnotesize
\begin{center}
\label{tab12} \caption{ Prediction accuracy comparison based on the mean (standard deviation) of leave one out cross-validation (LOCV) for all 200 data sets from random location samples. Smaller LOCV shows better performance.}  
\begin{tabular}{|l|l|l|l|l|l| }
 \hline 
 $\epsilon$ & n & MLR &LCM&  Envelope & Spatial Envelope   \\
 \hline 
1&100	&20.12 (1.613)	&21.01 (1.863)	&14.32 (1.699)	&14.98 (1.722)\\
&225	&19.34 (1.231)	&19.68 (1.542)	&13.12 (1.234)	& 13.19 (1.201)\\
&400	&17.83 (0.804)	&18.22 (1.101)	&11.73 (0.718)	& 12.37 (0.819)\\
\hline
2&100	&104.02 (36.702)	&23.32 (4.111)	&93.02 (30.433)	&19.21 (2.004)\\
&225	&102.41 (34.521)	&21.41 (3.758)	&91.34 (27.211)	&17.34 (1.352)\\
&400	&100.39 (30.822)	&19.20 (3.201)	&89.21 (25.581)	&16.68 (1.110)\\
\hline
3&100	&116.34 (45.089)	&25.21 (4.821)	&97.01 (43.021)	&20.79 (2.115)\\
&225	&108.15 (34.211)	&22.35 (3.555)	&95.52 (31.774)	&18.92 (1.944)\\
&400	&101.54 (32.102)	&20.44 (2.998)	&90.94 (30.234)	&17.03 (1.234)\\
\hline															
\end{tabular}
\end{center}}
\end{table}

From the summary of all three different error distributions, one can see that for the standard normal errors, where the observations are independent from each other, the spatial envelope provides comparable results to the envelope method and both performs better than MLR and LCM. In error distributions $2$ and $3$ where there exists spatial dependency in the data, the spatial envelope method performed almost equally as well as they did in the cases without spatial dependency while original envelope loses its efficiency. In addition, spatial envelope outperformed LCM in both independent and dependent cases. Since spatial envelope takes the spatial correlation among observations into consideration, it provides more accurate results compared to the original envelope model. Furthermore, spatial envelope only uses the material part of the data which leads to a more efficient results compared to LCM which uses both material and immaterial part of the data. Therefore, we can conclude that the proposed spatial envelope model provided consistent estimates with good prediction accuracy in all error distributions considered. This result is consistent for both sampling methods which indicates the spatial distribution of the observations has minimal impact on the estimation.   

As in \cite{cook2010envelope}, it is possible for an objective function defined on Grassmann manifolds to have multiple local optimal points. One way to check this is to run the simulation with different starting values and compare the results.

In order to investigate the accuracy of the asymptotic variance of $avar(\sqrt{n}\boldsymbol\beta^{*})$ that is presented in (\ref{asymbet}), we used the following simulation. The purpose of this simulation is to show that the variation of the spatial envelope estimator approaches its asymptotic variance derived in (\ref{asymbet}) when the sample size increases. The data is generated following model (\ref{simmod}) with five responses and one covariate i.e. $\textbf{Y}_i \in \mathbbm{R}^{5}$, $\textbf{X}_i \in \mathbbm{R}$, and the structural dimension $u=1$. In addition, we let $\boldsymbol\Omega_{1}=5\textbf{I}_{u}$, $\boldsymbol\Omega_{0}=\textbf{I}_{5-u}$ and $\boldsymbol\eta = 1$. 
The sample size $n$ is 100, 225, 400, and 900 , randomly chosen from a $101\times 101$ grid on $[0,1]^{2}$. For each sample size, 100 replications are performed to compute the estimation variance for the elements in $\hat{\boldsymbol\beta}$. For the spatial correlation, we used the  Matern covariance function with $\sigma_m=3$, $\theta_1=2$ and $\theta_2=0.5$.

\begin{figure}
    \centering
    \includegraphics[width=2.5in,height=6cm]{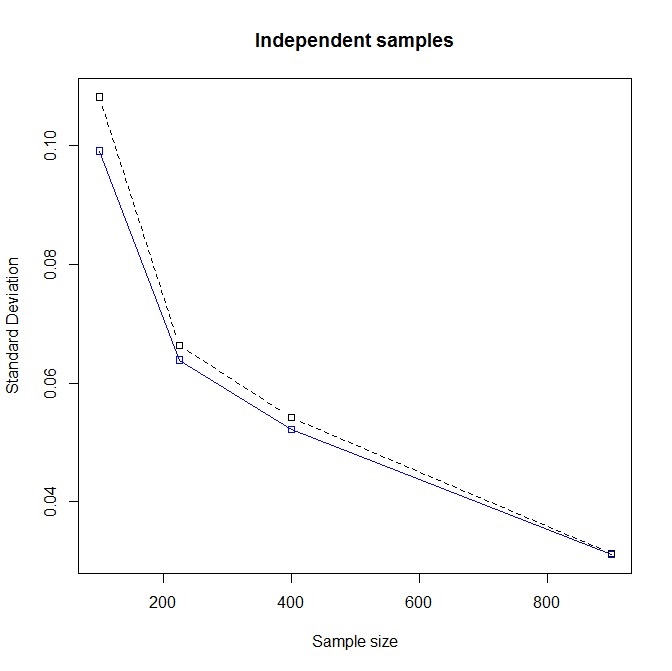}
    \includegraphics[width=2.5in,height=6cm]{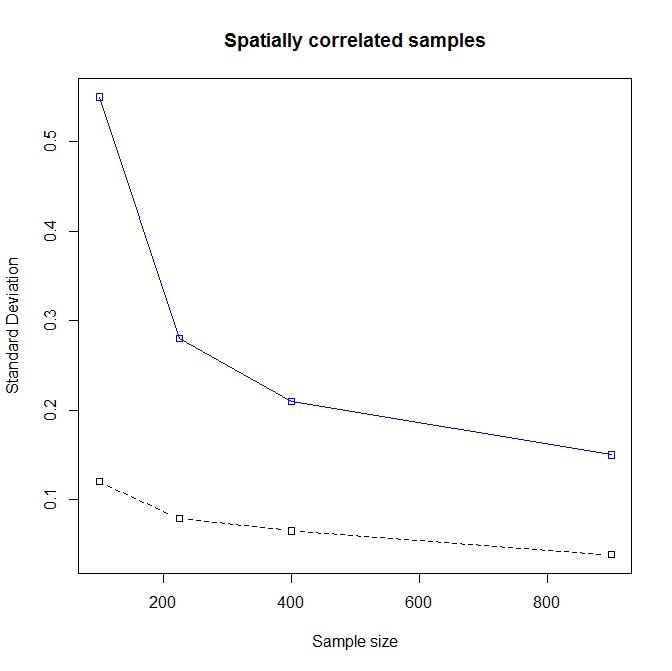}
    \caption{Simulation results of the asymptotic variance for a randomly selected element of $\hat{\boldsymbol\beta}$ for the envelope and the spatial envelope for the independent case (\textbf{left panel}) and for the spatially correlated data (\textbf{right panel}). The blue solid  line shows the estimated standard deviation of the envelope estimator and the black dash line denotes the estimated standard deviation of the spatial envelope estimator.}
    \label{asymgr}
\end{figure}

Figure \ref{asymgr} shows the simulation results of the asymptotic variance for a randomly selected element of $\hat{\boldsymbol\beta}$. The left panel of the figure \ref{asymgr} shows the asymptotic variance for the independent case and the right panel shows the same results for the spatially correlated data for the envelope and the spatial envelope. The blue line shows the estimated standard deviation of the envelope estimator and the black line denotes the estimated standard deviation of the spatial envelope estimator. From this figure, one can see that for the standard normal errors, where the observations are independent of each other, the variance of the spatial envelope and the envelope method are very similar. On the other hand, where there exists spatial dependency in the data, the spatial envelope method outperformed the envelope method. 


\section{Application}
In this section, we apply the proposed methodology to the air pollution data in the Northeastern United States. It is worth mentioning that the main purpose of this data analysis is to provide an insight that how the proposed approach can be used to find the reduced response space in multivariate spatial data analysis. This data has drawn much attention from both statisticians and scientists in other areas. Researchers looked at this data from different points of view including, but not restricted to, climate change \citep{phelan2016assessing}, health science  \citep{kioumourtzoglou2016long}, and air quality \citep{battye2016evaluating}. These studies showed that relationships exist between air pollution and meteorological factors, such as wind, temperature and humidity. Most of the existing studies focus on one of these pollutants, but since correlation exists among these pollutants, it is beneficial to study them simultaneously.

\begin{figure}
    \centering
    \includegraphics[width=2.5in,height=6cm]{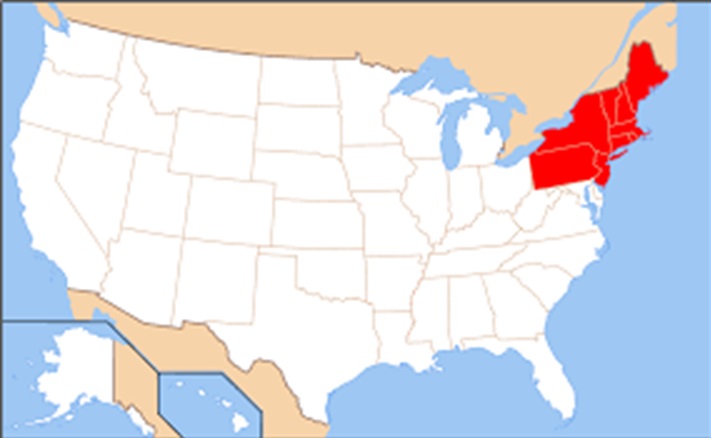}
    \includegraphics[width=2.5in,height=6cm]{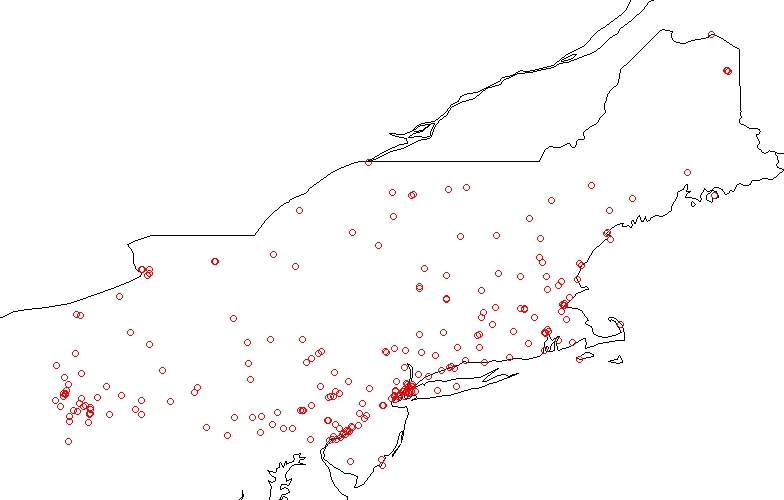}
    \caption{{\footnotesize
\textbf{Left}: Study area in the United States of America. States of interest are shaded in red. 
 \textbf{Right}: Location of different sites in the study area. It can be seen that there is a higher number of sites in places with larger population compare to other places in the study area.}}
    \label{fig1}
\end{figure}

The pollutants and weather data that we used in this study include the average levels of the following variables in January 2015. We choose a group of ambient air pollutants monitored by EPA because they present a high threat to human health. Specifically, we have 8 response variables: ground level ozone, sulfur dioxide ($SO_2$), carbon monoxide ($CO$), nitrogen dioxide ($NO_2$), nitrogen monoxide ($NO$), lead, PM 2.5, and PM 10. PM 10 includes particles less than or equal to 10 micrometers in diameter. Similarly, PM 2.5 includes particles less than or equal to 2.5 micrometers and is also called fine particle pollution. This data also includes the following meteorological variables: wind, temperature, and relative humidity as predictors. Along with this information, latitude and longitude of the monitoring locations are used to model the spatial structure in the data. Our study area consists of 9 states in the Northeast of the United States: Connecticut, Maine, Massachusetts, New Hampshire, New Jersey, New York, Pennsylvania, Rhode Island, and Vermont. This dataset  is available at \textcolor{blue}{http://aqsdr1.epa.gov/aqsweb/aqstmp/airdata/download\_files.html\#Daily}. Figure \ref{fig1} shows the study area and the location of 270 air monitoring sites. 

The preliminary analysis using Moran's I and plots of the empirical variogram determined that spatial correlation does exist in this data. The results of the preliminary analysis can be found in the Appendix, section 9.7. Cross-validation showed that the best choice for the structural dimension is 3. The Matern's covariance parameters, $\theta_{1}$ and $\theta_{2}$, are estimated to be 0.68 and 0.27, respectively. This estimates shows the existence of spatial dependency in the data. The corresponding direction estimates ($\hat{\boldsymbol\Gamma}_1$) from the spatial envelope are in Table 3. It is worth mentioning that the $\hat{\boldsymbol\Gamma}_1$ is not unique and it can be any orthonormal basis of the envelope subspace. The estimated regression coefficients and their standard deviation can be found in the Appendix, section 9.8.

\begin{table}[htp]
{\footnotesize
\begin{center}
\label{tab14} \caption{ The corresponding direction estimates using spatial envelope for the air pollution data in northeastern United States of America.}
\begin{tabular}{|l|l|l|l| }
 \hline 
Variable	&	Direction 1	&	Direction 2	&	Direction 3	\\
 \hline 
Ozone &-0.0464 & 0.0432  &   -0.0080\\
\hline 
Carbon monoxide & 0.2840  &   -0.3717   &  -0.0179\\
\hline 
Lead &   -0.0739 & 0.0872 & 0.0008\\
\hline 
Nitrogen dioxide &   -0.5089 & 0.2612  &   -0.4639\\
\hline 
Nitrogen monoxide &   -0.3056   &  -0.1137 & 0.2757\\
\hline 
Sulfur dioxide &  -0.5335 & 0.0241  &   -0.2981\\
\hline 
PM10 &   -0.3257   &  -0.8667  &   -0.0506\\
\hline 
PM2.5   &    -0.4106 & 0.1394 & 0.7855\\
\hline															
\end{tabular}
\end{center}}
\end{table}

\begin{figure}
    \centering
    \includegraphics[width=0.8\textwidth,height=6cm]{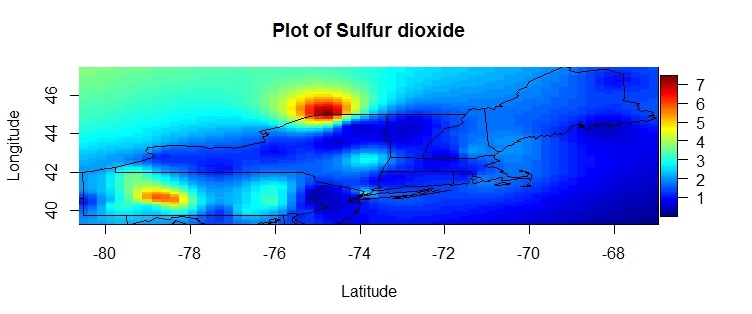}
    \caption{Prediction plot of the Sulfur dioxide for the study area. As it can be seen, the Sulfur dioxide is moderately high for the most part of the study area. Sulfur dioxide is extremely high in Johnstown where there exists a lot of defense manufacturing.}
    \label{fig5}
\end{figure}

 \begin{figure}
    \centering
    \includegraphics[width=0.8\textwidth,height=6cm]{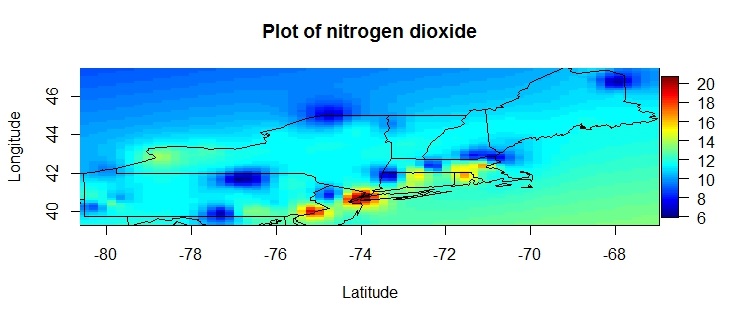}
    \caption{Prediction plot of the Nitrogen dioxide for the study area. The Nitrogen dioxide is high in Newark, New York, Philadelphia, and Rhodes Island which are all highly populated areas.}
    \label{fig6}
\end{figure}

\begin{figure}
    \centering
    \includegraphics[width=0.8\textwidth,height=6cm]{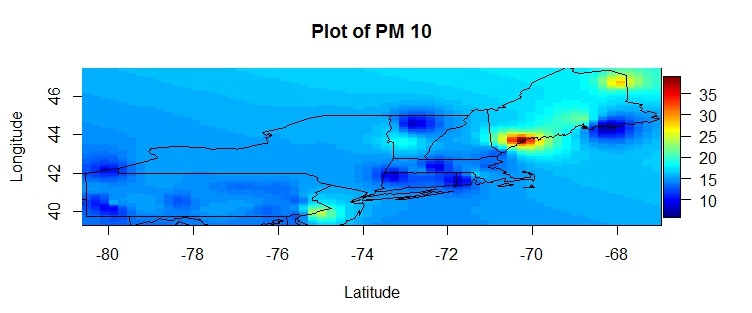}
    \caption{Prediction plot of the PM 10 for the study area. The PM 10 is high for most part of the study area especially in Philadelphia and Augusta.}
    \label{fig7}
\end{figure}

\begin{figure}
    \centering
    \includegraphics[width=0.8\textwidth,height=6cm]{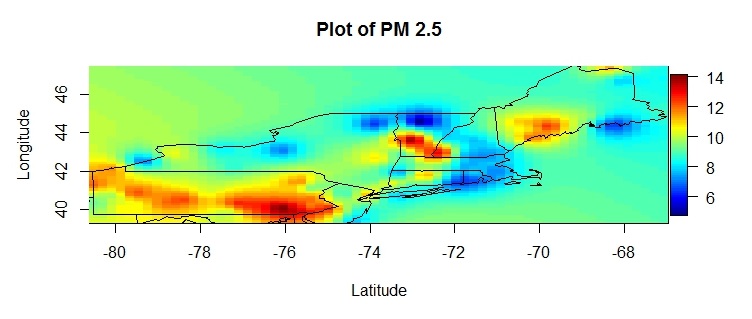}
    \caption{Prediction plot of the PM 2.5 for the study area. The PM 2.5 is moderately high in almost every place in the study area especially in Pennsylvania state, Augusta, and middle of Vermont state.}
    \label{fig9}
\end{figure}

By checking the estimated basis coefficients of the minimal subspace (directions) and the regression coefficients, we can see Sulfur dioxide, Nitrogen dioxide, PM 10, and PM 2.5 dominate each of the three directions, respectively. Using fossil fuels creates sulfur dioxide, nitrogen monoxide, and nitrogen dioxide. The nitrogen monoxide will also become nitrogen dioxide in the atmosphere. Existence of the particles in the air leads to reduction in visibility and causes the air to become hazy when levels are elevated. Furthermore, since these particles can travel deeply into the human lungs, they can cause health problem such as lung cancer. The main source of these particles in the air is from pollutants emitted from power plants, industries and automobiles. 

Figure \ref{fig5} to \ref{fig9} shows the prediction plots for the three pollutants with the largest impact. Figure \ref{fig5} shows the prediction plot of  the Sulfur dioxide for the study area. The Sulfur dioxide is moderately high for the most part of the study area. In addition, Sulfur dioxide is extremely high in Johnstown where there exists a lot of defense manufacturing. Figure \ref{fig6} shows the prediction plot of the Nitrogen dioxide for the study area. The Nitrogen dioxide  is high in Newark, New York, Philadelphia, and Rhodes Island which are all highly populated areas. Figure \ref{fig7} shows the prediction  plot of the PM 10 for the study area. The PM 10 is high for most part of the study area especially in Philadelphia and Augusta. Figure \ref{fig9} shows the prediction plot of the PM 2.5 for the study area. The PM 2.5 is moderately high in almost every place in the study area especially in Pennsylvania state, Augusta, and middle of Vermont state. Prediction plots of the other variables can be found in Appendix, section 9.9.

The square root of the leave one out cross-validation for MLR, LCM, envelope, and spatial envelope are 6.23, 7.05, 4.88, and 2.98, respectively. This result shows that spatial envelope outperforms other methods and provides more accurate prediction. In summary, we find out that the most important pollutants in January are particulates, sulfur, and nitrogen, and other pollutants have minimal effect. These statistical conclusions support the environmental chemical claim that in the cold weather, due to the fossil burning and inversion, sulfur dioxide, nitrogen dioxide, and particulate matters are the most important pollutants \citep{byers1959general, laegreid1999agriculture}.


\section{Conclusion}
Air pollution has a serious impact on human health. Research has greatly improved the understanding of each particular pollutant and their relationship with weather conditions. However, there are relatively few studies about the effects of meteorological variables on several pollutants together. Motivated
by an analysis of air pollution in the northeastern United States, we proposed a new parsimonious multivariate spatial model. Emphasis of this work is placed on inference and constructing a method that can provide more efficient estimation for the parameters of interest than traditional maximum likelihood estimators through capturing the spatial structure in the data.

Our model is flexible enough to characterize complex dependency and cross-dependency structures of different pollutants. From a simulation study and real data analysis, we showed that the proposed spatial envelope model outperforms multivariate linear regression, envelope, and  linear coregionalization models. This new approach provides more efficient estimation for regression coefficients compared to the traditional maximum likelihood approach. 

The method presented in this paper is for a multivariate spatial response with separable covariance matrix. This framework can be also extended to the cases that the covariance matrix is non-separable. Furthermore, current work assumes the normality in the derivations of the estimators. Confirming that the normality assumption is satisfied is more important for the spatial random fields than when working with envelope models. 
The violation of the normality assumption brings computational and theoretical challenges \cite{diggle1998model,liu2017spatial}. Incorporating the envelope idea with a multivariate non-Gaussian spatial random field, which is beyond the scope of this paper, is a very interesting and challenging topic. The mis-specification of the spatial structure  is also a very interesting and challenging topic. Investigation of the potential cost of mis-specifying the spatial correlation structure is also an interesting topic. The mis-specification can affect the estimation of the coefficient and prediction. Another possible extension of current methodology is for the case with spatiotemporal responses. The investigation for these more general cases is under way.



\begin{thebibliography}{}

\bibitem[\protect\citeauthoryear{abramowitz1964handbook}{Abramowitz and Stegun}{1964}]{abramowitz1964handbook}
Abramowitz, M. and Stegun, I. A. (1964). 
\newblock {\em Handbook of mathematical functions: with formulas, graphs, and mathematical tables}, Volume~{55}.
\newblock {Courier Corporation}.

\bibitem[\protect\citeauthoryear{battye2016evaluating}{Battye et al.}{2016}]{battye2016evaluating}
Battye, W. H., Bray, C. D., Aneja, V. P., Tong, D., Lee, P., and Tang, Y. (2016). 
\newblock Evaluating ammonia (NH 3) predictions in the NOAA National Air Quality Forecast Capability (NAQFC) using in situ aircraft, ground-level, and satellite measurements from the DISCOVER-AQ Colorado campaign. 
\newblock {\em Atmospheric Environment\/}~{\em 140}, pp. {342--351}.


\bibitem[\protect\citeauthoryear{byers1959general}{Byers}{1959}]{byers1959general}
Byers, H. R. (1959). 
\newblock {\em General meteorology}.
\newblock McGraw-Hill.


\bibitem[\protect\citeauthoryear{chiles2009geostatistics}{Chiles and Delfiner}{1999}]{chiles2009geostatistics}
Chiles, J P and Delfiner, P. (1999).
\newblock {\em Geostatistics: modeling spatial uncertainty}. Volume~{497}.
\newblock John Wiley \& Sons.

\bibitem[\protect\citeauthoryear{christensen2001advanced}{Christensen}{2001}]{christensen2001advanced}
Christensen, R. (2001). 
\newblock {\em Advanced linear modeling: multivariate, time series, and spatial data; nonparametric regression and response surface maximization}.
\newblock Springer Science \& Business Media.

\bibitem[\protect\citeauthoryear{cook2010envelope}{Cook, Li, and Chiaromonte}{2010}]{cook2010envelope}
Cook, R. D., Li, B., and Chiaromonte, F. (2010).
\newblock Envelope models for parsimonious and efficient multivariate linear regression.
\newblock {\em Statistica Sinica}, pp. {927--960}.


\bibitem[\protect\citeauthoryear{cook2015envlp}{Cook, Su, and Yang}{2014}]{cook2015envlp}
Cook, R. D., Su, Z. and Yang, Y. (2014). 
\newblock envlp: A MATLAB Toolbox for Computing Envelope Estimators in Multivariate Analysis.
\newblock {\em Journal of Statistical Software\/}~{62 (1)}, pp. {1--20}.


\bibitem[\protect\citeauthoryear{cook2015simultaneous}{Cook and Zhang}{2015}]{cook2015simultaneous}
Cook, R. D., and Zhang, X. (2015). 
\newblock Simultaneous envelopes for multivariate linear regression.
\newblock {\em Technometrics\/}~{57 (1)}, pp. {11--25}.

\bibitem[\protect\citeauthoryear{cook2016algorithms}{Cook and Zhang}{2016}]{cook2016algorithms}
Cook, R. D., and Zhang, X. (2016). 
\newblock Algorithms for envelope estimation. 
\newblock {\em Journal of Computational and Graphical Statistics\/}~{25 (1)}, pp. {284--300}.

\bibitem[\protect\citeauthoryear{cook2016note}{Cook, Forzani, and Su}{2016}]{cook2016note}
Cook, R. D., Forzani, L., and Su, Z. (2016).
\newblock A note on fast envelope estimation
\newblock {\em Journal of Multivariate Analysis\/}~{150}, pp. {42--54}.


\bibitem[\protect\citeauthoryear{cooper2003business}{Cooper, Schindler, and Sun}{2003}]{cooper2003business}
Cooper, D. R., Schindler, P. S., and Sun, J. (2003). 
\newblock {\em Business research methods}.
\newblock McGraw-Hill/Irwin New York, NY.

\bibitem[\protect\citeauthoryear{cressie2015statistics}{Cressie}{1993}]{cressie2015statistics}
Cressie, N. (1993). 
\newblock {\em Statistics for spatial data}.
\newblock John Wiley \& Sons.

\bibitem[\protect\citeauthoryear{diggle1998model}{Diggle, Tawn, and Moyeed}{1998}]{diggle1998model}
Diggle, P. J., Tawn, J. A., and Moyeed, R. A. (1998).
\newblock Model based geostatistics (with discussion). . 
\newblock {\em Journal of the Royal Statistical Society: Series C (Applied Statistics)\/}~{47 (3)}, pp. {299--350}.


\bibitem[\protect\citeauthoryear{edelman1998geometry}{Edelman, Arias, and Smith}{1998}]{edelman1998geometry}
Edelman, A., Arias, T. A., and Smith, S T. (1998).
\newblock The geometry of algorithms with orthogonality constraints. 
\newblock {\em SIAM journal on Matrix Analysis and Applications\/}~{20 (2)}, pp. {303--353}.

\bibitem[\protect\citeauthoryear{genton2015cross}{Genton and Kleiber}{2015}]{genton2015cross}
Genton, M. G., and Kleiber, W. (1991).
\newblock Cross-covariance functions for multivariate geostatistics.
\newblock {\em Statistical Science \/}~{30 (2)},, pp. {147--163}.

\bibitem[\protect\citeauthoryear{gneiting2010matern}{Gneiting, Kleiber, and Schlather}{2010}]{gneiting2010matern}
Gneiting, T., Kleiber, W., and Schlather, M. (2010).
\newblock Mat{\'e}rn cross-covariance functions for multivariate random fields.
\newblock {\em Journal of the American Statistical Association\/}~{105 (491)}, pp. {1167--1177}.


\bibitem[\protect\citeauthoryear{goulard1992linear}{Goulard and Voltz}{1992}]{goulard1992linear}
Goulard, M and Voltz, M. (1992).
\newblock Linear coregionalization model: tools for estimation and choice of cross-variogram matrix.
\newblock {\em Mathematical Geology\/}~{24 (3)}, pp. {269--286}.


\bibitem[\protect\citeauthoryear{guinness2014multivariate}{Guinness et al.}{2014}]{guinness2014multivariate}
Guinness, J., Fuentes, M., Hesterberg, D., and Polizzotto, M. (2014). 
\newblock Multivariate spatial modeling of conditional dependence in microscale soil elemental composition data.
\newblock {\em Spatial Statistics\/}~{9}, pp. {93--108}.

\bibitem[\protect\citeauthoryear{kioumourtzoglou2016long}{Kioumourtzoglou et al.}{2016}]{kioumourtzoglou2016long}
Kioumourtzoglou, M. A., Schwartz, J. D., Weisskopf, M. G., Melly, S. J., Wang, Y., Dominici, F., and Zanobetti, A. (2016). 
\newblock Long-term PM2. 5 exposure and neurological hospital admissions in the northeastern United States. 
\newblock {\em Environmental Health Perspectives (Online)\/}~{124 (11)}, pp.{23--29}.

\bibitem[\protect\citeauthoryear{laegreid1999agriculture}{L{\ae}greid, Bockman, and Kaarstad}{1999}]{laegreid1999agriculture}
L{\ae}greid, M and Bockman, O C and Kaarstad, O. (1999). 
\newblock {\em Agriculture, fertilizers and the environment.}
\newblock {CABI publishing}.

\bibitem[\protect\citeauthoryear{lave1973analysis}{Lave, and Seskin}{1973}]{lave1973analysis}
Lave, L. B., and Seskin, E. P. (1973). 
\newblock An analysis of the association between US mortality and air pollution.
\newblock {\em Journal of the American Statistical Association\/}~{68 (342)}, pp. {284--290}.

\bibitem[\protect\citeauthoryear{li2017parsimonious}{Li and Zhang}{2017}]{li2017parsimonious}
Li, L., and Zhang, X. (2017). 
\newblock Parsimonious tensor response regression.
\newblock {\em Journal of the American Statistical Association\/}, pp. {1--16}.

\bibitem[\protect\citeauthoryear{liang1992multivariate}{Liang, Zeger, and Qaqish}{1992}]{liang1992multivariate}
Liang, K. Y., Zeger, S. L., and Qaqish, B. (1992).
\newblock Multivariate regression analyses for categorical data.
\newblock {\em Journal of the Royal Statistical Society. Series B (Methodological)\/}, pp. {3--40}.

\bibitem[\protect\citeauthoryear{liu2017spatial}{Liu et al.}{1992}]{liu2017spatial}
Liu, X., Chen, F., Lu, Y. C., and Lu, C. T. (2017).
\newblock Prediction for Multivariate Non-Gaussian Data.
\newblock {\em ACM Transactions on Knowledge Discovery from Data (TKDD)\/},~ {11 (3)} pp. {36:1--36:27}.


\bibitem[\protect\citeauthoryear{myers1991pseudo}{Myers}{1991}]{myers1991pseudo}
Myers, D. E. (1991).
\newblock Pseudo-cross variograms, positive-definiteness, and cokriging.
\newblock {\em Mathematical Geology \/}~{23 (6)}, pp. {805--816}.

\bibitem[\protect\citeauthoryear{phelan2016assessing}{Phelan et al.}{2016}]{phelan2016assessing}
Phelan, J and Belyazid, S and Jones, P and Cajka, J and Buckley, J and Clark, C. (2016). 
\newblock Assessing the Effects of Climate Change and Air Pollution on Soil Properties and Plant Diversity in Sugar Maple--Beech--Yellow Birch Hardwood Forests in the Northeastern United States: Model Simulations from 1900 to 2100. 
\newblock {\em Water, Air, \& Soil Pollution\/}~{227 (3)}, pp. {1--30}.

\bibitem[\protect\citeauthoryear{seber2008matrix}{Seber}{2008}]{seber2008matrix}
Seber, G. A. F. (2008)
\newblock {\em A matrix handbook for statisticians}. Volume~{15}.
\newblock John Wiley \& Sons.

\bibitem[\protect\citeauthoryear{su2011partial}{Su and Cook}{2011}]{su2011partial}
Su, Z. and Cook, R. D. (2011). 
\newblock Partial envelopes for efficient estimation in multivariate linear regression.
\newblock {\em Biometrika\/}~{98(1)}, pp. {133--146}.

\bibitem[\protect\citeauthoryear{su2012inner}{Su and Cook}{2012}]{su2012inner}
Su, Z. and Cook, R. D. (2012). 
\newblock Inner envelopes: efficient estimation in multivariate linear regression.
\newblock {\em Biometrika\/}~{99 (3)}, pp. {687--702}.

\bibitem[\protect\citeauthoryear{su2013estimation}{Su and Cook}{2013}]{su2013estimation}
Su, Z. and Cook, R. D. (2013). 
\newblock Estimation of multivariate means with heteroscedastic errors using envelope models.
\newblock {\em Statistica Sinica}, pp.{213--230}.


\bibitem[\protect\citeauthoryear{ver1998constructing}{Ver Hoef and Barry}{1998}]{ver1998constructing}
Ver Hoef, J. M. and Barry, R. P. (1998). 
\newblock Constructing and fitting models for cokriging and multivariable spatial prediction.
\newblock {\em Journal of Statistical Planning and Inference\/}~{69 (2)},  pp. {275--294}.

\bibitem[\protect\citeauthoryear{wackernagel2013multivariate}{Wackernagel}{2003}]{wackernagel2013multivariate}
Wackernagel, H. (2003).
\newblock {\em Multivariate geostatistics: an introduction with applications}.
\newblock {Springer Science \& Business Media}


\bibitem[\protect\citeauthoryear{zhang2007maximum}{Zhang}{2007}]{zhang2007maximum}
Zhang, H. (2007).
\newblock Maximum-likelihood estimation for multivariate spatial linear coregionalization models,
\newblock {\em Environmetrics\/}, {18 (2)}, pp. {125--139}.


\bibitem[\protect\citeauthoryear{zhang2017tensor}{Zhang and Li}{2017}]{zhang2017tensor}
Zhang, X. and Li, L. (2017).
\newblock Tensor Envelope Partial Least-Squares Regression.
\newblock {\em Technometrics}, pp.{1--11}.

\bibitem[\protect\citeauthoryear{zhang2017model}{Zhang and Mai}{2017}]{zhang2017model}
Zhang, X. and Mai, Q. (2017)
\newblock {Model-free Envelope Dimension Selection}
\newblock {arXiv preprint arXiv:1709.03945}.

\bibitem[\protect\citeauthoryear{zhang2018functional}{Zhang, Wang, and Wu}{2018}]{zhang2018functional}
Zhang, X. Wang, C. and Wu, Y. (2018). 
\newblock {Functional envelope for model-free sufficient dimension reduction}.
\newblock {\em Journal of Multivariate Analysis}. {163}, pp. {37--50}.

\end{thebibliography}

\newpage
\begin{thebibliography}{}

\bibitem[\protect\citeauthoryear{cliff1973spatial}{Cliff and Ord}{1973}]{cliff1973spatial}
Cliff, A. D., and Ord, J. K. (1973). 
\newblock {\em Spatial Autocorrelation}.
\newblock Pion, London.

\bibitem[\protect\citeauthoryear{cook2010envelope}{Cook, Li, and Chiaromonte}{2010}]{cook2010envelope}
Cook, R. D., Li, B., and Chiaromonte, F. (2010).
\newblock Envelope models for parsimonious and efficient multivariate linear regression.
\newblock {\em Statistica Sinica}, pp. {927--960}.

\bibitem[\protect\citeauthoryear{goulard1992linear}{Goulard and Voltz}{1992}]{goulard1992linear}
Goulard, M and Voltz, M. (1992).
\newblock Linear coregionalization model: tools for estimation and choice of cross-variogram matrix.
\newblock {\em Mathematical Geology\/}~{24 (3)}, pp. {269--286}.


\bibitem[\protect\citeauthoryear{seber2008matrix}{Seber}{2008}]{seber2008matrix}
Seber, G. A. F. (2008)
\newblock {\em A matrix handbook for statisticians}. Volume~{15}.
\newblock John Wiley \& Sons.

\bibitem[\protect\citeauthoryear{shapiro1986as}{Shapiro}{1986}]{shapiro1986as}
Shapiro, A. (1986). 
\newblock Asymptotic theory of overparameterized structural models. 
\newblock {\em Journal of American Statistical Association}~ {81}, pp. {142--149}. 

\bibitem[\protect\citeauthoryear{zhang2007maximum}{Zhang}{2007}]{zhang2007maximum}
Zhang, H. (2007).
\newblock Maximum-likelihood estimation for multivariate spatial linear coregionalization models,
\newblock {\em Environmetrics\/}, {18 (2)}, pp. {125--139}.


\end{thebibliography}


\section{Appendix: Theoretical results and prediction plots}
\subsection{Derivation of the factorization of the likelihood function in section 4.1}

The likelihood function of the model (3.6) will be as follows:
\begin{equation}
\label{aslik2}
\begin{aligned}
&L^{u}(\boldsymbol\alpha, \boldsymbol\beta^{*},\textbf{V}_{0},\textbf{V}_{1},\boldsymbol\theta)=\left[\det((\textbf{V}_{0}+\textbf{V}_{1})\otimes \boldsymbol\rho(\boldsymbol\theta))\right]^{-\frac{1}{2}}\\
&\times exp\left\{-\frac{1}{2}(\mathbbm{Y}-\boldsymbol\alpha\otimes\textbf{1}_{n}- \mathbbm{X}\boldsymbol\beta^{*})^{T} \left((\textbf{V}_{0}+\textbf{V}_{1}) \otimes\boldsymbol\rho(\boldsymbol\theta)\right)^{-1}(\mathbbm{Y}-\boldsymbol\alpha\otimes\textbf{1}_{n}- \mathbbm{X}\boldsymbol\beta^{*})\right\}\\
&= [\det(\textbf{V}_{0} \boldsymbol\otimes \boldsymbol\rho(\boldsymbol\theta)+\textbf{V}_{1} \otimes\boldsymbol\rho(\boldsymbol\theta))]^{-\frac{1}{2}}\\
&\times exp\left\{-\frac{1}{2}(\mathbbm{Y}-\boldsymbol\alpha\otimes\textbf{1}_{n}-\mathbbm{X}\boldsymbol\beta^{*})^{T}\left((\textbf{V}_{0}+\textbf{V}_{1})^{-1}\otimes\boldsymbol\rho^{-1}(\boldsymbol\theta)\right)(\mathbbm{Y}-\boldsymbol\alpha\otimes\textbf{1}_{n}- \mathbbm{X}\boldsymbol\beta^{*})\right\}\\
&= [\det(\textbf{V}_{0}\otimes \boldsymbol\rho(\boldsymbol\theta)+\textbf{V}_{1}\otimes \boldsymbol\rho(\boldsymbol\theta))]^{-\frac{1}{2}}\\
&\times exp\left\{-\frac{1}{2}(\mathbbm{Y}-\boldsymbol\alpha\otimes\textbf{1}_{n}- \mathbbm{X}\boldsymbol\beta^{*})^{T} \left((\textbf{V}_{0}^{\dagger}\otimes\boldsymbol\rho^{-1}(\boldsymbol\theta))+(\textbf{V}_{1}^{\dagger}\otimes\boldsymbol\rho^{-1}(\boldsymbol\theta))\right)(\mathbbm{Y}-\boldsymbol\alpha\otimes\textbf{1}_{n}- \mathbbm{X}\boldsymbol\beta^{*})\right\},
\end{aligned}
\end{equation}
where $\dagger$ denotes Moore-Penrose inverse and $\textbf{V}_0=\boldsymbol\Gamma_{0}\boldsymbol\Omega_{0}\boldsymbol\Gamma_{0}$ and $\textbf{V}_1=\boldsymbol\Gamma_{1}\boldsymbol\Omega_{1}\boldsymbol\Gamma_{1}$. Since $span(\boldsymbol\beta) \subseteq span(\textbf{V}_{1})$ and $\boldsymbol\beta = \boldsymbol\Gamma_{1}\boldsymbol\eta$, therefore we have $\boldsymbol\beta^{T} = \boldsymbol\eta^{T}\boldsymbol\Gamma_{1}^{T}$ which means 
\[
\boldsymbol\beta^{*} = vec(\boldsymbol\beta^{T}) = vec(\boldsymbol\eta^{T}\boldsymbol\Gamma_{1}^{T}) = (\boldsymbol\Gamma_{1}\otimes\boldsymbol\eta^{T})vec(\textbf{I}_{u}).
\]
Last equality holds by the results of  theorem 11.6a in \cite{seber2008matrix}. Thus we have
\begin{equation*}
\begin{aligned}
(\textbf{V}_{0}^{\dagger}\otimes\boldsymbol\rho^{-1}(\boldsymbol\theta))\mathbbm{X}\boldsymbol\beta^{*} &= (\textbf{V}_{0}^{\dagger}\otimes\boldsymbol\rho^{-1}(\boldsymbol\theta))(\textbf{I}_{r}\otimes\textbf{X})\boldsymbol\beta^{*}\\
&= (\textbf{V}_{0}^{\dagger}\otimes\boldsymbol\rho^{-1}(\boldsymbol\theta))(\textbf{I}_{r}\otimes\textbf{X})(\boldsymbol\Gamma_{1}\otimes\boldsymbol\eta^{T})vec(\textbf{I}_{u})\\
& = (\textbf{V}_{0}^{\dagger}\boldsymbol\Gamma_{1}\otimes\boldsymbol\rho^{-1}(\boldsymbol\theta)\textbf{X}\boldsymbol\eta^{T})vec(\textbf{I}_{u})\\
& = (\boldsymbol\Gamma_{0}\boldsymbol\Omega_{0}^{-1}\boldsymbol\Gamma_{0}^{T} \boldsymbol\Gamma_{1}\otimes \boldsymbol\rho^{-1}(\boldsymbol\theta)\textbf{X} \boldsymbol\eta^{T})vec(\textbf{I}_{u})\\
& = \textbf{0},
\end{aligned}
\end{equation*}
the last equality holds because $\boldsymbol\Gamma_{1}$ and $\boldsymbol\Gamma_{0}$ are orthagonal. Therefore, Since $(\textbf{V}_{0}^{\dagger}\otimes\boldsymbol\rho^{-1}(\boldsymbol\theta))\mathbbm{X}\boldsymbol\beta^{*}= \textbf{0}$ and $\textbf{V}=\textbf{V}_0+\textbf{V}_1$, the likelihood in  (\ref{aslik2}) can be factored as: 
\begin{equation}
\label{aslik3}
\begin{aligned}
&L^{u}(\boldsymbol\alpha, \boldsymbol\beta^{*},\textbf{V}_{0},\textbf{V}_{1},\boldsymbol\theta)= [\det((\textbf{V}_{0}+\textbf{V}_{1})\otimes \boldsymbol\rho(\boldsymbol\theta)) \\
&\times exp\left\{-\frac{1}{2}(\mathbbm{Y}-\boldsymbol\alpha\otimes\textbf{1}_{n}- \mathbbm{X}\boldsymbol\beta^{*})^{T} \left(\textbf{V}_{1}^{\dagger}\otimes\boldsymbol\rho^{-1}(\boldsymbol\theta)\right) (\mathbbm{Y}-\boldsymbol\alpha\otimes\textbf{1}_{n}- \mathbbm{X}\boldsymbol\beta^{*})\right\}\\
&\times exp\left\{-\frac{1}{2}(\mathbbm{Y}-\boldsymbol\alpha\otimes\textbf{1}_{n})^{T} \left(\textbf{V}_{0}^{\dagger}\otimes\boldsymbol\rho^{-1}(\boldsymbol\theta)\right) (\mathbbm{Y}-\boldsymbol\alpha\otimes\textbf{1}_{n})\right\}\\
&=L_{1}^{u}(\boldsymbol\alpha, \boldsymbol\beta^{*},\textbf{V}_{1},\boldsymbol\theta)\times L_{2}^{u}(\boldsymbol\alpha, \textbf{V}_{0},\boldsymbol\theta),
\end{aligned}
\end{equation}
where
\begin{equation}
\label{aslik4}
\begin{aligned}
L_{1}^{u}(\boldsymbol\alpha, \boldsymbol\beta^{*},\textbf{V}_{1},\boldsymbol\theta)&= [det_{0}(\textbf{V}_{1})]^{-\frac{n}{2}}[\det(\boldsymbol\rho(\boldsymbol\theta))]^{-\frac{r}{2}} \\
&\times exp\left\{-\frac{1}{2}(\mathbbm{Y}-\boldsymbol\alpha\otimes\textbf{1}_{n}- \mathbbm{X}\boldsymbol\beta^{*})^{T} \left(\textbf{V}_{1}^{\dagger}\otimes\boldsymbol\rho^{-1}(\boldsymbol\theta)\right) (\mathbbm{Y}-\boldsymbol\alpha\otimes\textbf{1}_{n}- \mathbbm{X}\boldsymbol\beta^{*})\right\},\\
L_{2}^{u}(\boldsymbol\alpha, \textbf{V}_{0},\boldsymbol\theta) & = [det_{0}(\textbf{V}_{0})]^{-\frac{n}{2}}[\det(\boldsymbol\rho(\boldsymbol\theta))]^{-\frac{r}{2}}\\
&\times exp\left\{-\frac{1}{2}(\mathbbm{Y}-\boldsymbol\alpha\otimes\textbf{1}_{n})^{T} \left(\textbf{V}_{0}^{\dagger}\otimes\boldsymbol\rho^{-1}(\boldsymbol\theta)\right) (\mathbbm{Y}-\boldsymbol\alpha\otimes\textbf{1}_{n})\right\},
\end{aligned}
\end{equation}
where $det_{0}(\textbf{A})$ denotes the product of non-zero eigenvalues of \textbf{A} where \textbf{A} is a non-zero symmetric matrix. This is due to
\begin{equation*}
\begin{aligned}
\det((\textbf{V}_{0}+\textbf{V}_{1})\otimes \boldsymbol\rho(\boldsymbol\theta)) &= \det[\textbf{V}_{0}\otimes \boldsymbol\rho(\boldsymbol\theta)+\textbf{V}_{1}\otimes \boldsymbol\rho(\boldsymbol\theta)]\\
&= det_{0}[\textbf{V}_{0}\otimes \boldsymbol\rho(\boldsymbol\theta)] +  det_{0}[\textbf{V}_{1}\otimes \boldsymbol\rho(\boldsymbol\theta)] \\
&= [det_{0}(\textbf{V}_{0})]^{n}[det_{0}(\boldsymbol\rho(\boldsymbol\theta))]^{r} + [det_{0}(\textbf{V}_{1})]^{n}[det_{0}(\boldsymbol\rho(\boldsymbol\theta))]^{r}\\
&= [det_{0}(\textbf{V}_{0})]^{n}[\det(\boldsymbol\rho(\boldsymbol\theta))]^{r} + [det_{0}(\textbf{V}_{1})]^{n}[\det(\boldsymbol\rho(\boldsymbol\theta))]^{r}
\end{aligned}
\end{equation*}
the last equality holds because is $\boldsymbol\rho(\boldsymbol\theta)$ a full rank positive definite matrix therefore $det_{0}=\det$.


\subsection{Coordinate free version of the algorithm of the spatial envelope}
The objective is to maximize the likelihood in (3.7) over $\boldsymbol\alpha, \boldsymbol\beta^{*},\textbf{V}_{0},\textbf{V}_{1}$, and $\boldsymbol\theta$ subject to the constraints:
\begin{equation}
\begin{aligned}
\label{con}
\begin{split}
&span(\boldsymbol\beta) \subseteq span(\textbf{V}_{1}),~~ (a)\\
&\textbf{V}_{0}\textbf{V}_{1} = 0,~~~~~~~~~~~~~~~~~ (b).
\end{split}
\end{aligned}
\end{equation}
Based on this factorization given in equation (\ref{aslik3}), we can decompose the likelihood maximization into the following steps:
\begin{enumerate}
\item Fix $\boldsymbol\beta,\textbf{V}_{0}$, $\textbf{V}_{1}$, and $\boldsymbol\theta$, and maximize $L^{(u)}$ in (3.6) over $\boldsymbol\alpha$ which will be:
\[
\hat{\boldsymbol\alpha} = \bar{\textbf{Y}}-\bar{\textbf{X}}\boldsymbol\beta^{T}.
\]
Let $\textbf{H}=\textbf{Y}-\bar{\textbf{Y}}\otimes\textbf{1}_{n}$, $\textbf{U}=vec(\textbf{H})$, $\textbf{G} = \textbf{X}-\bar{\textbf{X}}\otimes \textbf{1}_{n}$, and $\textbf{F}=\textbf{I}_{r}\otimes\textbf{G}$. Therefore, the profile likelihood can be written as the following:
\begin{equation}
\label{plik1}
\begin{aligned}
L_{1}^{u}(\boldsymbol\beta^{*},\textbf{V}_{1},\boldsymbol\theta)&= [det_{0}(\textbf{V}_{1})]^{-\frac{n}{2}}[\det(\boldsymbol\rho(\boldsymbol\theta))]^{-\frac{r}{2}} \\
&\times exp\left\{-\frac{1}{2}(\textbf{U}- \textbf{F}\boldsymbol\beta^{*})^{T} \left(\textbf{V}_{1}^{\dagger}\otimes\boldsymbol\rho^{-1}(\boldsymbol\theta)\right) (\textbf{U}-\textbf{F}\boldsymbol\beta^{*})\right\},
\end{aligned}
\end{equation}
and
\begin{equation}
\label{plik2}
L_{2}^{u}(\textbf{V}_{0},\boldsymbol\theta)  = [det_{0}(\textbf{V}_{0})]^{-\frac{n}{2}} [\det(\boldsymbol\rho(\boldsymbol\theta))]^{-\frac{r}{2}} exp\left\{-\frac{1}{2}\textbf{U}^{T} \left(\textbf{V}_{0}^{\dagger}\otimes\boldsymbol\rho^{-1}(\boldsymbol\theta)\right) \textbf{U}\right\}.
\end{equation}

\item Fix $\textbf{V}_{1}$, and $\boldsymbol\theta$ and maximize the function $L_{1}^{u}$ over $\boldsymbol\beta^{*}$, subject to (\ref{con}a), to obtain $L_{21}^{u}(\textbf{V}_{1},\boldsymbol\theta)$. Since $vec(\textbf{A}\textbf{B})=(\textbf{I}_{r}\otimes\textbf{A})vec(\textbf{B}^{T})$ and 
\[
tr(\textbf{D}^{T}(\textbf{C}^{T}\textbf{B}^{T}\textbf{A}^{T})) = (vec(\textbf{D}))^{T}(\textbf{A}\otimes\textbf{C}^{T})(vec(\textbf{B}))^{T},
\] 
we have 
\begin{equation}
\label{step2}
\begin{aligned}
(\textbf{U}- \textbf{F}\boldsymbol\beta^{*})^{T} &\left(\textbf{V}_{1}^{\dagger}\otimes\boldsymbol\rho^{-1}(\boldsymbol\theta)\right) (\textbf{U}-\textbf{F}\boldsymbol\beta^{*}) = tr\left((\textbf{H}-\textbf{G}\boldsymbol\beta^{T})^{T}\boldsymbol\rho^{-1}(\boldsymbol\theta)(\textbf{H}-\textbf{G}\boldsymbol\beta^{T})\textbf{V}_{1}^{\dagger}\right)\\
&= tr\left((\textbf{H}-\textbf{G}\boldsymbol\beta^{T})^{T}\boldsymbol\rho^{-\frac{1}{2}}(\boldsymbol\theta)\boldsymbol\rho^{-\frac{1}{2}}(\boldsymbol\theta) (\textbf{H}-\textbf{G}\boldsymbol\beta^{T})\textbf{V}_{1}^{\dagger}\right)\\
&= tr\left(\boldsymbol\rho^{-\frac{1}{2}}(\boldsymbol\theta)(\textbf{H}-\textbf{G}\boldsymbol\beta^{T})\textbf{V}_{1}^{\dagger}(\textbf{H}-\textbf{G}\boldsymbol\beta^{T})^{T}\boldsymbol\rho^{-\frac{1}{2}}(\boldsymbol\theta)\right)\\
&= tr\left(\left(\boldsymbol\rho^{-\frac{1}{2}}(\boldsymbol\theta)\textbf{H}-\boldsymbol\rho^{-\frac{1}{2}}(\boldsymbol\theta)\textbf{G}\boldsymbol\beta^{T}\right)\textbf{V}_{1}^{\dagger}\left(\boldsymbol\rho^{-\frac{1}{2}}(\boldsymbol\theta)\textbf{H}-\boldsymbol\rho^{-\frac{1}{2}}(\boldsymbol\theta)\textbf{G}\boldsymbol\beta^{T}\right)^{T}\right)\\
&= tr\left(\left(\boldsymbol\rho^{-\frac{1}{2}}(\boldsymbol\theta)\textbf{H}-\boldsymbol\rho^{-\frac{1}{2}}(\boldsymbol\theta)\textbf{G}\boldsymbol\beta^{T}\textbf{I}_{r}\right)\textbf{V}_{1}^{\dagger}\left(\boldsymbol\rho^{-\frac{1}{2}}(\boldsymbol\theta)\textbf{H}-\boldsymbol\rho^{-\frac{1}{2}}(\boldsymbol\theta)\textbf{G}\boldsymbol\beta^{T}\textbf{I}_{r}\right)^{T}\right)
\end{aligned}
\end{equation}
where $tr(\cdot)$ denotes the trace of the matrix. The last equality in equation (\ref{step2}) is from Lemma 4.1 in \cite{cook2010envelope}. Thus, the optimal $\boldsymbol\rho^{-\frac{1}{2}}(\boldsymbol\theta) \textbf{G}\boldsymbol\beta^{T}\textbf{I}_{r}$ is 
\[
\textbf{P}_{\left(\boldsymbol\rho^{-\frac{1}{2}}(\boldsymbol\theta) \textbf{G}\right)} \left(\boldsymbol\rho^{-\frac{1}{2}}(\boldsymbol\theta) \textbf{H}\right)\textbf{P}_{(\textbf{I}_{r}(\textbf{V}_{1}^{\dagger}))}^{T}= \textbf{P}_{\left(\boldsymbol\rho^{-\frac{1}{2}}(\boldsymbol\theta)\textbf{G}\right)} \left(\boldsymbol\rho(\boldsymbol\theta)^{-\frac{1}{2}} \textbf{H}\right)\textbf{P}_{\textbf{V}_{1}}, 
\]
where $\textbf{P}_{(\cdot)}$ is the projection onto the subspace indicated by its argument. This implies following
\[
\boldsymbol\beta^{T}=\left(\textbf{G}^{T}\boldsymbol\rho^{-1}(\boldsymbol\theta)\textbf{G}\right)^{-1}\textbf{G}\boldsymbol\rho^{-1}(\boldsymbol\theta)
\textbf{H}\textbf{P}_{\textbf{V}_{1}}\Rightarrow \boldsymbol\beta = \textbf{P}_{\textbf{V}_{1}}\hat{\boldsymbol\beta},
\]
where $\boldsymbol\beta$ is the MLE estimate of $\boldsymbol\beta$ from the full model (3.6). Substituting this into (\ref{plik2}) and using the relation $\textbf{P}_{\textbf{V}_{1}}\textbf{V}_{1}^{\dagger}=\textbf{V}_{1}^{\dagger}$, the maximum of $L_{2}^{(u)}$ for fixed $\textbf{V}_{1}$ over $\boldsymbol\beta$ is
\begin{footnotesize}
\begin{equation}
\label{l11}
\begin{aligned}
&L_{11}^{u}(\textbf{V}_{1},\boldsymbol\theta)=[det_{0}(\textbf{V}_{1})]^{-\frac{n}{2}}[\det(\boldsymbol\rho(\boldsymbol\theta))]^{-\frac{r}{2}}\\  
&\times exp\left\{-\frac{1}{2}tr\left(\left(\boldsymbol\rho(\boldsymbol\theta)^{-\frac{1}{2}}\textbf{H}-\textbf{P}_{\left(\boldsymbol\rho^{-\frac{1}{2}}(\boldsymbol\theta) \textbf{G}\right)}\boldsymbol\rho^{-\frac{1}{2}}(\boldsymbol\theta)\textbf{H}\textbf{P}_{\textbf{V}_{1}}\right)\textbf{V}_{1}^{\dagger} \left(\boldsymbol\rho(\boldsymbol\theta)^{-\frac{1}{2}}\textbf{H}-\textbf{P}_{\left(\boldsymbol\rho^{-\frac{1}{2}}(\boldsymbol\theta) \textbf{G}\right)}\boldsymbol\rho^{-\frac{1}{2}}(\boldsymbol\theta)\textbf{H}\textbf{P}_{\textbf{V}_{1}}\right)^{T}\right)\right\}\\
&=[det_{0}(\textbf{V}_{1})]^{-\frac{n}{2}}[\det(\boldsymbol\rho(\boldsymbol\theta))]^{-\frac{r}{2}}\\  
&\times exp\left\{-\frac{1}{2}tr\left(\left(\boldsymbol\rho^{-\frac{1}{2}}(\boldsymbol\theta)\textbf{H}-\textbf{P}_{\left(\boldsymbol\rho^{-\frac{1}{2}}(\boldsymbol\theta)\textbf{G}\right)}\boldsymbol\rho^{-\frac{1}{2}}(\boldsymbol\theta)\textbf{H}\right)\textbf{V}_{1}^{\dagger} \left(\boldsymbol\rho^{-\frac{1}{2}}(\boldsymbol\theta)\textbf{H}-\textbf{P}_{\left(\boldsymbol\rho^{-\frac{1}{2}}(\boldsymbol\theta)\textbf{G}\right)}\boldsymbol\rho^{-\frac{1}{2}}(\boldsymbol\theta)\textbf{H}\right)^{T}\right)\right\}\\
&=[det_{0}(\textbf{V}_{1})]^{-\frac{n}{2}}[\det(\boldsymbol\rho(\boldsymbol\theta))]^{-\frac{r}{2}} exp\left\{-\frac{1}{2}tr\left(\left(\textbf{Q}_{\left(\boldsymbol\rho^{-\frac{1}{2}}(\boldsymbol\theta) \textbf{G}\right)}\boldsymbol\rho^{-\frac{1}{2}}(\boldsymbol\theta)\textbf{H}\right) \textbf{V}_{1}^{\dagger}\left(\textbf{Q}_{\left(\boldsymbol\rho^{-\frac{1}{2}}(\boldsymbol\theta) \textbf{G}\right)}\boldsymbol\rho^{-\frac{1}{2}}(\boldsymbol\theta)\textbf{H}\right)^{T}\right)\right\}
\end{aligned}
\end{equation}
\end{footnotesize}
where $\textbf{Q}_{\left(\boldsymbol\rho^{-\frac{1}{2}}(\boldsymbol\theta) \textbf{G}\right)} = \textbf{I}_n - \textbf{P}_{\left(\boldsymbol\rho^{-\frac{1}{2}}(\boldsymbol\theta) \textbf{G}\right)}$.

\item Maximize $L^{u}(\textbf{V}_{0},\textbf{V}_{1},\boldsymbol\theta)$ over all $\textbf{V}_{0}$, $\textbf{V}_{1}$, and $\boldsymbol\theta$. Since $L^{u}(\textbf{V}_{0},\textbf{V}_{1},\boldsymbol\theta) = L_{1}^{u}(\textbf{V}_{1},\boldsymbol\theta)\times L_{2}^{u}(\textbf{V}_{0},\boldsymbol\theta)$, we have
\begin{equation}
\begin{aligned}
&L^{u}(\textbf{V}_{0},\textbf{V}_{1},\boldsymbol\theta) = [det_{0}(\textbf{V}_{0})]^{-\frac{n}{2}}[det_{0}(\textbf{V}_{1})]^{-\frac{n}{2}}[\det(\boldsymbol\rho(\boldsymbol\theta))]^{-r}\\
&\times exp\left\{-\frac{1}{2}tr\left(\left(\textbf{Q}_{\left(\boldsymbol\rho^{-\frac{1}{2}}(\boldsymbol\theta) \textbf{G}\right)}\boldsymbol\rho^{-\frac{1}{2}}(\boldsymbol\theta)\textbf{H}\right) \textbf{V}_{1}^{\dagger}\left(\textbf{Q}_{\left(\boldsymbol\rho^{-\frac{1}{2}}(\boldsymbol\theta) \textbf{G}\right)}\boldsymbol\rho^{-\frac{1}{2}}(\boldsymbol\theta)\textbf{H}\right)^{T}\right)\right\}\\
&\times exp\left\{-\frac{1}{2}\textbf{U}^{T} \left(\textbf{V}_{0}^{\dagger}\otimes\boldsymbol\rho^{-1}(\boldsymbol\theta)\right) \textbf{U}\right\}\\
&=[det_{0}(\textbf{V}_{0})]^{-\frac{n}{2}}[det_{0}(\textbf{V}_{1})]^{-\frac{n}{2}}[\det(\boldsymbol\rho(\boldsymbol\theta))]^{-r}\\
&\times exp\left\{-\frac{1}{2}tr\left(\left(\textbf{Q}_{\left(\boldsymbol\rho^{-\frac{1}{2}}(\boldsymbol\theta) \textbf{G}\right)}\boldsymbol\rho^{-\frac{1}{2}}(\boldsymbol\theta)\textbf{H}\right) \textbf{V}_{1}^{\dagger}\left(\textbf{Q}_{\left(\boldsymbol\rho^{-\frac{1}{2}}(\boldsymbol\theta) \textbf{G}\right)}\boldsymbol\rho^{-\frac{1}{2}}(\boldsymbol\theta)\textbf{H}\right)^{T}\right)\right\}\\
&\times exp\left\{-\frac{1}{2}tr\left( \boldsymbol\rho^{-\frac{1}{2}}(\boldsymbol\theta)\textbf{H} \textbf{V}_{0}^{\dagger} \textbf{H}^{T} \boldsymbol\rho^{-\frac{1}{2}}(\boldsymbol\theta)\right)\right\}.
\end{aligned}
\end{equation}
This maximization can be as follows:
\begin{enumerate}
     \item Fix $\textbf{V}_{0}$ and $\textbf{V}_{1}$ and maximize $L^{u}(\textbf{V}_{0},\textbf{V}_{1},\boldsymbol\theta)$ over $\boldsymbol\theta$ by solving the following maximization problem:
\begin{small}
\begin{equation*}
\begin{aligned}
\hat{\boldsymbol\theta}&=\argmaxF_{\boldsymbol\theta} \{r \det(\boldsymbol\rho(\boldsymbol\theta))\\
&+\frac{1}{2}tr\left(\left( \textbf{Q}_{\left(\boldsymbol\rho^{-\frac{1}{2}}(\boldsymbol\theta)\textbf{G}\right)}\boldsymbol\rho^{-\frac{1}{2}}(\boldsymbol\theta)\textbf{H}\right) \textbf{V}_{1}^{\dagger}\left(\textbf{Q}_{\left(\boldsymbol\rho^{-\frac{1}{2}}(\boldsymbol\theta)\textbf{G}\right)}\boldsymbol\rho(\boldsymbol\theta)^{-\frac{1}{2}}\textbf{H}\right)^{T} + \boldsymbol\rho^{-\frac{1}{2}}(\boldsymbol\theta)\textbf{H} \textbf{V}_{0}^{\dagger} \textbf{H}^{T} \boldsymbol\rho^{-\frac{1}{2}}(\boldsymbol\theta)\right)\}.
\end{aligned}
\end{equation*}
\end{small}
     \item Fix the $\boldsymbol\theta$ and maximize $L^{u}(\textbf{V}_{0},\textbf{V}_{1},\boldsymbol\theta)$ over $\textbf{V}_{0}$ and $\textbf{V}_{1}$. This means maximize $L_{11}^{u}(\textbf{V}_{1},\boldsymbol\theta)$ over $\textbf{V}_{1}$ and $L_{12}^{u}(\textbf{V}_{0},\boldsymbol\theta)$ over $\textbf{V}_{0}$. Maximization $L_{11}^{u}(\textbf{P}_{\textbf{V}_{1}})$ over $\textbf{V}_{1}$ is 
\begin{equation}
L_{11}^{u}(\textbf{P}_{\textbf{V}_{1}}) \propto \left[ det_{0} \left(\textbf{P}_{\textbf{V}_{1}}\left( \textbf{H}^{T} \boldsymbol\rho^{-\frac{1}{2}}(\boldsymbol\theta)\textbf{Q}_{\left(\boldsymbol\rho^{-\frac{1}{2}}(\boldsymbol\theta)\textbf{G}\right)} \boldsymbol\rho^{-\frac{1}{2}}(\boldsymbol\theta)\textbf{H}\right)\textbf{P}_{\textbf{V}_{1}} \right)\right]^{-\frac{n}{2}}
\end{equation} 
and maximization $L_{12}^{u}(\textbf{P}_{\textbf{V}_{0}})$ over $\textbf{V}_{0}$ is
\begin{equation}
L_{12}^{u}(\textbf{P}_{\textbf{V}_{0}}) \propto \left[ det_{0} \left(\textbf{P}_{\textbf{V}_{0}} \textbf{H}^{T} \boldsymbol\rho^{-1}(\boldsymbol\theta)\textbf{H}\textbf{P}_{\textbf{V}_{0}} \right)\right]^{-\frac{n}{2}}.
\end{equation} 
Therefore, maximization $L^{u}(\textbf{V}_{0},\textbf{V}_{1},\boldsymbol\theta)$ over $\textbf{V}_{0}$ and $\textbf{V}_{1}$ is equivalent to maximization of $L_{11}^{u}(\textbf{P}_{\textbf{V}_{1}}) \times L_{12}^{u}(\textbf{P}_{\textbf{V}_{0}})$ which is proportion to 
\begin{small}
\begin{equation}
\label{logd}
\begin{aligned}
\textbf{D}&=\left[ det_{0}\left(\textbf{P}_{\textbf{V}_{1}}\left( \textbf{H}^{T} \boldsymbol\rho^{-\frac{1}{2}}(\boldsymbol\theta)\textbf{Q}_{\left(\boldsymbol\rho^{-\frac{1}{2}}(\boldsymbol\theta)\textbf{G}\right)} \boldsymbol\rho^{-\frac{1}{2}}(\boldsymbol\theta)\textbf{H}\right)\textbf{P}_{\textbf{V}_{1}} \right)\right]^{-\frac{n}{2}}\\
& \times \left[ det_{0} \left(\textbf{P}_{\textbf{V}_{0}} \textbf{H}^{T} \boldsymbol\rho^{-1}(\boldsymbol\theta)\textbf{H}\textbf{P}_{\textbf{V}_{0}} \right)\right]^{-\frac{n}{2}}\\
&= \left[det_{0} \left(\textbf{P}_{\textbf{V}_{1}}\left( \textbf{H}^{T} \boldsymbol\rho^{-\frac{1}{2}}(\boldsymbol\theta)\textbf{Q}_{\left(\boldsymbol\rho^{-\frac{1}{2}}(\boldsymbol\theta)\textbf{G}\right)} \boldsymbol\rho^{-\frac{1}{2}}(\boldsymbol\theta)\textbf{H}\right)\textbf{P}_{\textbf{V}_{1}}+\textbf{P}_{\textbf{V}_{0}} \textbf{H}^{T} \boldsymbol\rho^{-1}(\boldsymbol\theta)\textbf{H}\textbf{P}_{\textbf{V}_{0}} \right)\right]^{-\frac{n}{2}}\\
&= \left[det_{0} \left(\textbf{P}_{\textbf{V}_{1}}\left( \textbf{H}^{T} \boldsymbol\rho^{-\frac{1}{2}}(\boldsymbol\theta)\textbf{Q}_{\left(\boldsymbol\rho^{-\frac{1}{2}}(\boldsymbol\theta)\textbf{G}\right)} \boldsymbol\rho^{-\frac{1}{2}}(\boldsymbol\theta)\textbf{H}\right)\textbf{P}_{\textbf{V}_{1}}+\textbf{Q}_{\textbf{V}_{0}} \textbf{H}^{T} \boldsymbol\rho^{-1}(\boldsymbol\theta)\textbf{H}\textbf{Q}_{\textbf{V}_{0}} \right)\right]^{-\frac{n}{2}}
\end{aligned}
\end{equation}
\end{small} 
where $\textbf{Q}_{\textbf{V}_{0}} = \textbf{I}_{r}-\textbf{P}_{\textbf{V}_{1}}$. Since $\hat{\boldsymbol\Sigma}_{\textbf{Y}} = \textbf{H}^{T} \boldsymbol\rho^{-1}(\boldsymbol\theta)\textbf{H}$ and 
\begin{equation}
\begin{aligned}
\hat{\boldsymbol\Sigma}_{res} &= \textbf{H}^{T} \boldsymbol\rho^{-\frac{1}{2}}(\boldsymbol\theta)\textbf{Q}_{\left(\boldsymbol\rho^{-\frac{1}{2}}(\boldsymbol\theta)\textbf{G}\right)} \boldsymbol\rho^{-\frac{1}{2}}(\boldsymbol\theta)\textbf{H}\\ 
&=\textbf{H}^{T} \boldsymbol\rho^{-1}(\boldsymbol\theta)\textbf{H} \\
&- \textbf{H}^{T} \boldsymbol\rho^{-1}(\boldsymbol\theta)\textbf{G} \left(  \textbf{G}^{T} \boldsymbol\rho^{-1}(\boldsymbol\theta)\textbf{G} \right)^{-1}\textbf{G}^{T} \boldsymbol\rho^{-1}(\boldsymbol\theta)\textbf{H}.
\end{aligned}
\end{equation}
Therefore we have $\textbf{D} =  det(\textbf{P}_{\textbf{V}_{1}}\hat{\boldsymbol\Sigma}_{\textbf{res}}\textbf{P}_{\textbf{V}_{1}}+ \textbf{Q}_{\textbf{V}_{1}}\hat{\boldsymbol\Sigma}_{\textbf{Y}}\textbf{Q}_{\textbf{V}_{1}})$ and $\hat{\textbf{V}_{1}} = \argmaxF_{\textbf{V}_{1}} (\textbf{D})$ and $\textbf{P}_{\hat{\textbf{V}_{0}}} = \textbf{I}_{r} - \textbf{P}_{\hat{\textbf{V}_{1}}}$
\end{enumerate}
Repeat (a) and (b) until the difference between estimations of the parameters from two consecutive iterations is smaller than a pre-specified tolerance level.
\end{enumerate}


\subsection{Proof of Lemma 1}

In this section, we derive the Fisher information matrix for the parameters given by equation (4.2). Before starting the derivation, the following properties hold:

\begin{enumerate}
\item Suppose \textbf{A} and \textbf{X} are both $r \times r$, and 
\textbf{X} is symmetric, then
\begin{align*}
\frac{\partial vech(\textbf{X}^{-1})}{(\partial vech(\textbf{X}))^{T}} &= -\textbf{C}_{r}\left(\textbf{X}^{-1}\otimes \textbf{X}^{-1}\right)\textbf{E}_{r},
\end{align*}
where $\textbf{E}_{r}\in R^{r^{2}\times r(r+1)/2}$ is an expansion matrix such that for a matrix $\textbf{A}$, $vec(\textbf{A}) = \textbf{E}_{r}vech(\textbf{A})$, and $\textbf{C}_{r}\in R^{ r(r+1)/2\times r^{2}}$ is expansion matrix which is defined such that for a given matrix such as $\textbf{A}$, $vech(\textbf{A}) = \textbf{C}_{r}vec(\textbf{A})$ and $\textbf{E}_{r}\in R^{r^{2}\times r(r+1)/2}$ is expansion matrix which is defined such that $vec(\textbf{A}) = \textbf{E}_{r}vech(\textbf{A})$.
\item If \textbf{Y} = \textbf{A}\textbf{X}\textbf{B}, then 
\[
tr(\textbf{Y}) = vec(\textbf{A}^{T}\textbf{B}^{T})vec(X)= vec(\textbf{A}^{T}\textbf{B}^{T})\textbf{E}_{n}vech(X),
\]
and 
\[
\frac{\partial tr(\textbf{Y})}{\partial vec(\textbf{X})} = vec(\textbf{A}^{T}\textbf{B}^{T}).
\]
\item Suppose $\textbf{B}_1$ is an $m \times n$ and $\textbf{B}_2$ is an $n \times q$, matrix, then
\[
vec(\textbf{B}_1\textbf{B}_2) = (\textbf{B}_2\otimes \textbf{I}_{m})vec(\textbf{B}_1).
\]
\item Suppose $\textbf{X}$ is an $m \times n$ and $\textbf{A}$ is an $n \times n$, matrix, then
\[
\frac{\partial vec(\textbf{X}\textbf{A}\textbf{X})}{\partial (vec(\textbf{X}))^{T}} = (\textbf{X}^{T}\textbf{A}^{T}\otimes \textbf{I}_{n})\textbf{I}_{nm} + (\textbf{I} _{n}\otimes \textbf{X}^{T}\textbf{A}).
\]
\item Assume \textbf{X} to be $m\times n$ . Then we have, 
\[
\frac{\partial (\textbf{X}^{T}\textbf{AX})}{\partial \textbf{X}}= \textbf{AX}+\textbf{A}^{T}\textbf{X}.
\]

\item Let $\textbf{P}_{\textbf{E}_r}$ denotes the projection of $\textbf{E}_r(\textbf{E}^{T}_r\textbf{E}_r)^{-1}\textbf{E}^{T}_r$ then, $\textbf{P}_{\textbf{E}_r}=\textbf{E}_r\textbf{C}_r$ and $\textbf{E}^{T}_r\textbf{E}_r\textbf{C}_r=\textbf{E}^{T}_r$, 
\end{enumerate}
Proof of the first five properties can be found in \cite{seber2008matrix}. The proof of the last property can be found in \cite{cook2010envelope}

The logarithm of the likelihood function (3.7) is 
\begin{equation}
\label{loglik1}
\ell(\boldsymbol\Theta) = -\frac{1}{2}\log[\det(\textbf{V}\otimes \boldsymbol\rho(\boldsymbol\theta))]-\frac{1}{2}(\mathbbm{Y}-\boldsymbol\alpha\otimes \textbf{1}_{n}-\mathbbm{X}\boldsymbol\beta^{*})^{T}(\textbf{V}\otimes \boldsymbol\rho(\boldsymbol\theta))^{-1}(\mathbbm{Y}-\boldsymbol\alpha\otimes \textbf{1}_{n}-\mathbbm{X}\boldsymbol\beta^{*})
\end{equation}
where $\boldsymbol\Theta = \{\textbf{V}, \boldsymbol\alpha,\boldsymbol\beta^{*}, \boldsymbol\theta \}$. First and second derivatives of the log likelihood function in (\ref{loglik1}) with respect to $\boldsymbol\beta^{*}$ are
\begin{equation*}
\label{loglik2}
\begin{aligned}
\text{First derivative:}~\frac{\partial \ell(\boldsymbol\Theta)}{\partial \boldsymbol\beta^{*}} &= \mathbbm{X}^{T}(\textbf{V}^{-1}\otimes \boldsymbol\rho^{-1}(\boldsymbol\theta)) (\mathbbm{Y}-\boldsymbol\alpha\otimes \textbf{1}_{n}-\mathbbm{X}\boldsymbol\beta^{*}), \\
\text{Second derivative:}~\frac{\partial^{2} \ell(\boldsymbol\Theta)}{\partial \boldsymbol\beta^{*}\partial {\boldsymbol\beta^{*}}^{T}} &= -\mathbbm{X}^{T}(\textbf{V}^{-1}\otimes \boldsymbol\rho^{-1}(\boldsymbol\theta)) \mathbbm{X}\\
& = -(\textbf{I}_{r}\otimes \textbf{X}^{T})(\textbf{V}^{-1}\otimes \boldsymbol\rho^{-1}(\boldsymbol\theta))(\textbf{I}_{r}\otimes \textbf{X})\\
&=- \textbf{V}^{-1}\otimes \left(\textbf{X}^{T}\boldsymbol\rho^{-1}(\boldsymbol\theta)) \textbf{X}\right)
\end{aligned}
\end{equation*}

From (3.7), we can rewrite the log likelihood function as
\begin{equation}
\label{loglik5}
\begin{aligned}
\ell(\boldsymbol\Theta) &= -\frac{n}{2}\log[\det(\textbf{V})]-\frac{r}{2}\log [\det(\boldsymbol\rho(\boldsymbol\theta))]\\
&-\frac{1}{2}tr\left(\left(\boldsymbol\rho^{-\frac{1}{2}}(\boldsymbol\theta)\textbf{H}-\boldsymbol\rho^{-\frac{1}{2}}(\boldsymbol\theta)\textbf{G}\boldsymbol\beta^{T}\right)\textbf{V}^{-1} \left(\boldsymbol\rho^{-\frac{1}{2}}(\boldsymbol\theta)\textbf{H}-\boldsymbol\rho^{-\frac{1}{2}}(\boldsymbol\theta)\textbf{G}\boldsymbol\beta^{T}\right)^{T} \right).
\end{aligned}
\end{equation}
The $tr(\cdot)$ is due to 
\begin{equation*}
\label{loglik6}
\begin{aligned}
&(\textbf{U}-\textbf{F}\boldsymbol\beta^{*})^{T}(\textbf{V}^{-1}\otimes \boldsymbol\rho^{-1}(\boldsymbol\theta))(\textbf{U}-\textbf{F}\boldsymbol\beta^{*})= tr\left((\textbf{H}-\textbf{G}\boldsymbol\beta^{T})^{T} \boldsymbol\rho^{-1}(\boldsymbol\theta) (\textbf{H}-\textbf{G}\boldsymbol\beta^{T})^{T}\textbf{V}^{-1} \right)\\
&= tr\left(\left(\boldsymbol\rho^{-\frac{1}{2}}(\boldsymbol\theta)\textbf{H}-\boldsymbol\rho^{-\frac{1}{2}}(\boldsymbol\theta)\textbf{G}\boldsymbol\beta^{T}\right)\textbf{V}^{-1} \left(\boldsymbol\rho^{-\frac{1}{2}}(\boldsymbol\theta)\textbf{H}-\boldsymbol\rho^{-\frac{1}{2}}(\boldsymbol\theta)\textbf{G}\boldsymbol\beta^{T}\right)\right).
\end{aligned}
\end{equation*}
Therefore, the first derivative of the log likelihood function in (\ref{loglik5}) with respect to $\textbf{V}$ is
$\frac{\partial \ell(\boldsymbol\Theta)}{\partial vech(\textbf{V})} =\frac{\partial \ell(\boldsymbol\Theta)}{\partial vec(\textbf{V})}\frac{\partial vec(\textbf{V})}{\partial vech(\textbf{V})}$, where
\begin{equation}
\label{loglik6a}
\begin{aligned}
\frac{\partial \ell(\boldsymbol\Theta)}{\partial vech(\textbf{V})} &= -\frac{n}{2} vec\left( \textbf{V}^{-1}\right)^T \textbf{E}_r \\
& + \frac{1}{2}vec\left\{\textbf{V}^{-1}\left(\boldsymbol\rho^{-\frac{1}{2}}(\boldsymbol\theta)\textbf{H}-\boldsymbol\rho^{-\frac{1}{2}}(\boldsymbol\theta)\textbf{G}\boldsymbol\beta^{T}\right)^{T} \left(\boldsymbol\rho^{-\frac{1}{2}}(\boldsymbol\theta)\textbf{H}-\boldsymbol\rho^{-\frac{1}{2}}(\boldsymbol\theta)\textbf{G}\boldsymbol\beta^{T}\right)\textbf{V}^{-1}\right\}\textbf{E}_{r}\\
&= -\frac{n}{2} vech\left( \textbf{V}^{-1} \right)^T \textbf{E}^T_r\textbf{E}_r \\
& + \frac{1}{2}vech\left\{ \textbf{V}^{-1}\left(\boldsymbol\rho^{-\frac{1}{2}}(\boldsymbol\theta)\textbf{H}-\boldsymbol\rho^{-\frac{1}{2}}(\boldsymbol\theta)\textbf{G}\boldsymbol\beta^{T}\right)^{T} \left(\boldsymbol\rho^{-\frac{1}{2}}(\boldsymbol\theta)\textbf{H}-\boldsymbol\rho^{-\frac{1}{2}}(\boldsymbol\theta)\textbf{G}\boldsymbol\beta^{T}\right)\textbf{V}^{-1}\right\}\textbf{E}_{r}^{T}\textbf{E}_{r}\\
\end{aligned}
\end{equation}
and second derivative of the log likelihood function in (\ref{loglik5}) with respect to $\textbf{V}$ is
\begin{small}
\begin{eqnarray}
\label{loglik6b}
\frac{\partial^{2} \ell(\boldsymbol\Theta)}{\partial vech(\textbf{V})\partial vech(\textbf{V})^{T}} &=& \frac{n}{2} \textbf{E}^T_{r}\left(\textbf{V}^{-1} \otimes \textbf{V}^{-1} \right)\textbf{E}_{r}\cr
&-&\frac{1}{2} \textbf{A}\textbf{V}^{-1}\textbf{E}_{r}^{T}(\textbf{V}^{-1}\otimes \textbf{V}^{-1})\textbf{C}_{r}^{T}\textbf{E}_{r}^{T}\textbf{E}_{r} - \frac{1}{2} \textbf{A}^{T}\textbf{V}^{-1}\textbf{E}_{r}(\textbf{V}^{-1}\otimes \textbf{V}^{-1})\textbf{C}_{r}^{T}\textbf{E}_{r}^{T}\textbf{E}_{r}~~~~~~~~
\end{eqnarray}
\end{small}
where $ \textbf{A} = \left(\boldsymbol\rho^{-\frac{1}{2}}(\boldsymbol\theta)\textbf{H}-\boldsymbol\rho^{-\frac{1}{2}}(\boldsymbol\theta)\textbf{G}\boldsymbol\beta^{T}\right)^{T} \left(\boldsymbol\rho^{-\frac{1}{2}}(\boldsymbol\theta)\textbf{H}-\boldsymbol\rho^{-\frac{1}{2}}(\boldsymbol\theta)\textbf{G}\boldsymbol\beta^{T}\right)$. Thus, 
\[
E\left(\frac{\partial^{2} \ell(\boldsymbol\Theta)}{\partial vech(\textbf{V})\partial vech(\textbf{V})^{T}} \right) = -\frac{n}{2} \textbf{E}_{r}^{T}(\textbf{V}^{-1}\otimes \textbf{V}^{-1})\textbf{E}_{r}
\]

Finally, we have to calculate $\frac{\partial^{2} \ell(\boldsymbol\Theta)}{\partial \partial \boldsymbol\beta^{*} \partial vech(\textbf{V})^{T} }$ and $\frac{\partial^{2} \ell(\boldsymbol\Theta)}{\partial vech(\textbf{V})\partial \boldsymbol\beta^{*^{T} }}$. Since these two are equal, we only calculate the second one. 
\begin{equation}
\label{loglik7}
\begin{aligned}
&\frac{\partial^{2} \ell(\boldsymbol\Theta)}{\partial vech(\textbf{V})\partial \boldsymbol\beta^{*^{T}} } = \frac{\partial^{2} \ell(\boldsymbol\Theta)}{\partial vech(\textbf{V})\partial (vec(\boldsymbol\beta^{T}))^{T} }\\ 
&= \frac{1}{2} \frac{vec\left\{ \textbf{V}^{-1}\left(\boldsymbol\rho^{-\frac{1}{2}}(\boldsymbol\theta)\textbf{H}-\boldsymbol\rho^{-\frac{1}{2}}(\boldsymbol\theta)\textbf{G}\boldsymbol\beta^{T}\right)^{T} \left(\boldsymbol\rho^{-\frac{1}{2}}(\boldsymbol\theta)\textbf{H}-\boldsymbol\rho^{-\frac{1}{2}}(\boldsymbol\theta)\textbf{G}\boldsymbol\beta^{T}\right)\textbf{V}^{-1}\right\}\textbf{E}_{r}}{\partial (vec(\boldsymbol\beta^{T}))^{T} }\\
&= \frac{1}{2}\frac{vec\left[\textbf{V}^{-1}\left(\textbf{H}^{T}\boldsymbol\rho^{-1}(\boldsymbol\theta)\textbf{H} - \boldsymbol\beta\textbf{G}^{T}\boldsymbol\rho^{-1}(\boldsymbol\theta)\textbf{H} - \textbf{H}^{T}\boldsymbol\rho^{-1}(\boldsymbol\theta)\textbf{G}\boldsymbol\beta^{T} + \boldsymbol\beta\textbf{G}^{T}\boldsymbol\rho^{-1}(\boldsymbol\theta)\textbf{G}\boldsymbol\beta^{T} \right)\textbf{V}^{-1}\right]\textbf{E}_{r}}{\partial (vec(\boldsymbol\beta^{T}))^{T} }.
\end{aligned}
\end{equation}
The derivative of $vec\left(\textbf{V}^{-1}\textbf{H}^{T}\boldsymbol\rho^{-1}(\boldsymbol\theta)\textbf{H}\textbf{V}^{-1}\right)\textbf{E}_{r}$ with respect to $vec(\boldsymbol\beta^{T}))^{T}$ is zero. Furthermore, using matrix algebra, we have 
\begin{equation*}
\begin{aligned}
vec\left(\textbf{V}^{-1}\boldsymbol\beta\textbf{G}^{T}\boldsymbol\rho^{-1}(\boldsymbol\theta)\textbf{H}\textbf{V}^{-1}\right) &= \left(\textbf{V}^{-1}\textbf{H}^{T}\boldsymbol\rho^{-1}\textbf{G}\otimes \textbf{V}^{-1} \right)vec(\boldsymbol\beta)\\
&= \left(\textbf{V}^{-1}\textbf{H}^{T}\boldsymbol\rho^{-1}\textbf{G}\otimes \textbf{V}^{-1} \right)\textbf{K}_{rp}vec(\boldsymbol\beta^{T})   \\
vec\left(\textbf{V}^{-1}\textbf{H}^{T}\boldsymbol\rho^{-1}(\boldsymbol\theta)\textbf{G}\boldsymbol\beta^{T}\textbf{V}^{-1} \right) &=  \left(\textbf{V}^{-1}\otimes \textbf{V}^{-1}\textbf{H}^{T}\boldsymbol\rho^{-1}(\boldsymbol\theta)\textbf{G} \right)vec(\boldsymbol\beta^{T}).
\end{aligned}
\end{equation*}
where $\textbf{K}_{rp}\in \mathbbm{R}^{rp\times rp}$ is the unique matrix that transform the $vec$ of a matrix into the $vec$ of its transpose i.e. for a given matrix such as $\textbf{A}\in \mathbbm{R}^{m\times n}$ we have $vec(\textbf{A}^{T}) = \textbf{K}_{mn}vec(\textbf{A})$. More properties of  $\textbf{K}_{mn}$ can be found in \cite{cook2010envelope} lemma D.2. Therefore, we have 
\begin{equation}
\label{loglik88}
\begin{aligned}
\frac{vec\left(\textbf{V}^{-1}\boldsymbol\beta\textbf{G}^{T}\boldsymbol\rho^{-1}(\boldsymbol\theta)\textbf{H}\textbf{V}^{-1}\right)}{\partial (vec(\boldsymbol\beta^{T}))^{T} } &= \left(\textbf{V}^{-1}\textbf{H}^{T}\boldsymbol\rho^{-1}\textbf{G}\otimes \textbf{V}^{-1} \right)\textbf{K}_{rp}  \\
\frac{vec\left(\textbf{V}^{-1}\textbf{H}^{T}\boldsymbol\rho^{-1}(\boldsymbol\theta)\textbf{G}\boldsymbol\beta^{T} \textbf{V}^{-1}\right)}{\partial (vec(\boldsymbol\beta^{T}))^{T} } &= \left(\textbf{V}^{-1}\otimes\textbf{V}^{-1}\textbf{H}^{T}\boldsymbol\rho^{-1}(\boldsymbol\theta)\textbf{G} \right)\\
\frac{vec\left(\textbf{V}^{-1}\boldsymbol\beta\textbf{G}^{T}\boldsymbol\rho^{-1}(\boldsymbol\theta)\textbf{G}\boldsymbol\beta^{T} \textbf{V}^{-1}\right)}{\partial (vec(\boldsymbol\beta^{T}))^{T} } &=  \left(\textbf{V}^{-1}\boldsymbol\beta\textbf{G}^{T}\boldsymbol\rho^{-1}(\boldsymbol\theta)\textbf{G}\otimes\textbf{V}^{-1} \right)\textbf{K}_{rp} + \left(\textbf{V}^{-1}\otimes \textbf{V}^{-1}\boldsymbol\beta\textbf{G}^{T}\boldsymbol\rho^{-1}(\boldsymbol\theta)\textbf{G}\right).
\end{aligned}
\end{equation}
Substituting (\ref{loglik88}) in equation (\ref{loglik7}), we have 
\begin{equation}
\label{loglik71}
\begin{aligned}
\frac{\partial^{2} \ell(\boldsymbol\Theta)}{\partial vech(\textbf{V})\partial \boldsymbol\beta^{*^{T} }} &= \frac{1}{2}\left\{\textbf{V}^{-1}\left(\textbf{H}-\textbf{G}\boldsymbol\beta^{T}\right)^{T} \boldsymbol\rho^{-1}(\boldsymbol\theta)\textbf{G} \otimes \textbf{V}^{-1}\right\}\textbf{K}_{rp}\textbf{E}_{r} \\
&+ \frac{1}{2}\left\{\textbf{V}^{-1}\otimes\textbf{V}^{-1}\left(\textbf{H}-\textbf{G}\boldsymbol\beta^{T} \right)^{T} \boldsymbol\rho^{-1}(\boldsymbol\theta)\textbf{G}\right\}\textbf{E}_{r} 
\end{aligned}
\end{equation}

Taking the expected value of these derivatives together and the fact that 
\[
E\left[ \frac{\partial^{2} \ell(\boldsymbol\Theta)}{\partial vech(\textbf{V})\partial \boldsymbol\beta^{*} } \right]= \textbf{0},
\]
lead to obtain (4.4).


\subsection{Proof of Theorem 1}

In this section, we derive the an explicit expression for $\boldsymbol\Psi$ as given by (4.3). In order to find these expression, we need to find expressions for the eight partial derivatives $\frac{\partial \Psi_i}{\partial \phi_{j}^{T}}$ for $i = 1, 2$ and $j = 1, 2, 3, 4$.

\textbf{Theorem 1:} Suppose $\bar{\textbf{X}} = 0$ and $\textbf{J}$ is the Fisher information for $\psi(\phi)$ in the model (3.6):
\begin{equation*}
\begin{aligned}
\textbf{J}&= \begin{bmatrix}
\frac{1}{n}\mathbbm{X}^{T}\left( \textbf{V}^{-1}\otimes\boldsymbol\rho^{-1}(\boldsymbol\theta)\right)\mathbbm{X} & \textbf{0}\\
\textbf{0} &  \frac{1}{2}\textbf{E}^T_{r}\left(\textbf{V}^{-1}\otimes\textbf{V}^{-1}\right)\textbf{E}_{r}\\
\end{bmatrix}
\\
&=  
\begin{bmatrix}
\textbf{V}^{-1} \otimes \left(\frac{\textbf{X}^{T}\boldsymbol\rho^{-1}(\boldsymbol\theta)\textbf{X}}{n}\right) & \textbf{0}\\
\textbf{0} &  \frac{1}{2}\textbf{E}^T_{r}\left(\textbf{V}^{-1}\otimes\textbf{V}^{-1}\right)\textbf{E}_{r}\\
\end{bmatrix}.
\end{aligned}
\end{equation*}
Then
\begin{equation}
\label{asym5}
\sqrt{n}(\hat{\phi}-\phi)\rightarrow N(\textbf{0},\boldsymbol\Lambda_{0})
\end{equation}
where $\boldsymbol\Lambda_{0} = \Psi(\Psi^{T}\boldsymbol\Lambda\Psi)^{\dagger}\Psi$, $\boldsymbol\Lambda = \textbf{J}^{-1}$ is the asymptotic variance of the MLE under the full model, and $\Psi$ is as follows:
\begin{equation*}
\begin{bmatrix}
\textbf{K}_{rp} (\textbf{I}_{p}\otimes\boldsymbol\Gamma_{1}) &  \textbf{K}_{rp} (\boldsymbol\eta^{T}\otimes\textbf{I}_{r})& \textbf{0}& \textbf{0}\\\textbf{0} &  2\textbf{C}_{r}(\boldsymbol\Gamma_{1}\boldsymbol\Omega_{1}\otimes\textbf{I}_{r}-\boldsymbol\Gamma_{1}\otimes\boldsymbol\Gamma_{0}\boldsymbol\Omega_{0}\boldsymbol\Gamma_{0}^{T}) &  \textbf{C}_{r}(\boldsymbol\Gamma_{1}\otimes\boldsymbol\Gamma_{1})\textbf{E}_{u} & \textbf{C}_{r}(\boldsymbol\Gamma_{0}\otimes\boldsymbol\Gamma_{0})\textbf{E}_{r-u} \\
\end{bmatrix}.
\end{equation*}
Furthermore, $\boldsymbol\Lambda^{-\frac{1}{2}}(\boldsymbol\Lambda-\boldsymbol\Lambda_{0})\boldsymbol\Lambda^{-\frac{1}{2}} \geq 0$, so the spatial envelope model decreases the asymptotic variance.

\textbf{Proof}: We can rewrite $\boldsymbol\beta^{*}$ as follows
\begin{equation}
\label{loglik8}
\begin{aligned}
\boldsymbol\beta^{*} &= vec(\boldsymbol\eta^{T}\boldsymbol\Gamma_{1}^{T})\\
&= \textbf{K}_{rp}vec(\boldsymbol\Gamma_{1}\boldsymbol\eta)\\
& = \textbf{K}_{rp} (\textbf{I}_{p}\otimes\boldsymbol\Gamma_{1})vec(\boldsymbol\eta)\\
& = \textbf{K}_{rp} (\boldsymbol\eta^{T}\otimes\textbf{I}_{r})vec(\boldsymbol\Gamma_{1}).
\end{aligned}
\end{equation}

Therefore, the derivatives of $\psi_{1}$ with respect to $\phi_{1}^{T}$ is  
\begin{equation*}
\frac{\partial \psi_{1}}{\partial \phi_{1}^{T}} =\frac{\partial \boldsymbol\beta^{*}}{\partial (vec(\boldsymbol\eta))^{T}} = \frac{\partial \left[\textbf{K}_{rp} (\textbf{I}_{p}\otimes\boldsymbol\Gamma_{1})vec(\boldsymbol\eta)\right]}{\partial (vec(\boldsymbol\eta))^{T}}  = \textbf{K}_{rp} (\textbf{I}_{p}\otimes\boldsymbol\Gamma_{1}), 
\end{equation*}
and the derivatives of $\psi_{1}$ with respect to $\phi_{2}^{T}$ is 
\begin{equation}
\frac{\partial \psi_{1}}{\partial \phi_{2}^{T}} =\frac{\partial \boldsymbol\beta^{*}}{\partial (vec(\boldsymbol\Gamma))^{T}} = \frac{\partial  \left[\textbf{K}_{rp} (\boldsymbol\eta^{T}\otimes\textbf{I}_{r})vec(\boldsymbol\Gamma_{1})\right]}{\partial (vec(\boldsymbol\Gamma_{1}))^{T}} = \textbf{K}_{rp} (\boldsymbol\eta^{T}\otimes\textbf{I}_{r}).
\end{equation}
 It is clear that $\frac{\partial \psi_{1}}{\partial \phi_{3}^{T}} =\frac{\partial \psi_{1}}{\partial \phi_{4}^{T}} = \textbf{0}$. 

The derivative of $\frac{\partial \psi_{2}}{\partial \phi_{1}^{T}}$ to $\frac{\partial \psi_{2}}{\partial \phi_{4}^{T}}$ are similar to those in \cite{cook2010envelope}.  Having these derivatives together lead to obtain (4.3).

The asymptotic distribution (\ref{asym5}) follows from \cite{shapiro1986as}. In order to prove that $\boldsymbol\Lambda_{0} \leq \boldsymbol\Lambda$, we have 
\[
\boldsymbol\Lambda_{0} - \boldsymbol\Lambda = \textbf{J}^{-1} - \Psi(\Psi^{T}\boldsymbol\Lambda\Psi)^{\dagger}\Psi
= \textbf{J}^{-\frac{1}{2}}\left[ \textbf{I}_{pr+r(r+1)/2} - 
\textbf{J}^{\frac{1}{2}}\Psi(\Psi^{T}\boldsymbol\Lambda\Psi)^{\dagger}\Psi\textbf{J}^{\frac{1}{2}}
\right]\textbf{J}^{-\frac{1}{2}}
\]
Since the matrix $\textbf{I}_{pr+r(r+1)/2} - \textbf{J}^{\frac{1}{2}}\Psi(\Psi^{T}\boldsymbol\Lambda\Psi)^{\dagger}\Psi\textbf{J}^{\frac{1}{2}}$  is the projection on to orthogonal complement of $span(\textbf{J}^{\frac{1}{2}}\Psi)$, it is positive semidefinite, which implies that $\boldsymbol\Lambda_{0} - \boldsymbol\Lambda$ is also positive semidefinite. In addition, we have 
\[
\boldsymbol\Lambda^{-\frac{1}{2}}(\boldsymbol\Lambda-\boldsymbol\Lambda_{0})\boldsymbol\Lambda^{-\frac{1}{2}} = 
\textbf{I}_{pr+r(r+1)/2} - \textbf{J}^{\frac{1}{2}}\Psi(\Psi^{T}\boldsymbol\Lambda\Psi)^{\dagger}\Psi\textbf{J}^{\frac{1}{2}}
\]
which proves the last statement of the theorem.


\subsection{Proof of Corollary 1}

In this section, we restate and proof the corollary 1. 

\textbf{Corollary 1:} The asymptotic variance (avar) of $\sqrt{n}\boldsymbol\beta^*$ can be written as
\begin{equation}
\label{asymp7}
avar(\sqrt{n}\boldsymbol\beta^*) = \textbf{K}_{rp}\left\{\left(\frac{\textbf{X}^{T}\boldsymbol\rho(\boldsymbol\theta)^{-1}\textbf{X}}{n}\right)^{-1}\otimes \boldsymbol\Gamma_{1}\boldsymbol\Omega_{1}\boldsymbol\Gamma_{1}^{T} + (\boldsymbol\eta^{T}\otimes \boldsymbol\Gamma_{0})(\Psi_{2}^{T}\textbf{J}\Psi_{2})^{\dagger}(\boldsymbol\eta\otimes \boldsymbol\Gamma_{0}^{T})\right\}\textbf{K}_{rp}^T
\end{equation}
where $\Psi_{2} = \left(\frac{\partial \psi_{1}}{\partial  \phi_{2}^{T}}, \frac{\partial \psi_{2}}{\partial  \phi_{2}^{T}}\right)^{T}$. 

\textbf{Proof}: Using lemma 1 and theorem 1, the asymptotic variance of $\sqrt{n}\boldsymbol\beta^*$ can be written as
\[
avar(\sqrt{n}\boldsymbol\beta^*) = K_{1}(\Psi_{1}^{T}\textbf{J}\Psi_{1})^{\dagger}K_{1}^{T} + K_{2}(\Psi_{2}^{T}\textbf{J}\Psi_{2})^{\dagger}K_{2}^{T}
\]
where $\Psi_{1} = \left(\frac{\partial \psi_{1}}{\partial  \phi_{1}^{T}}, \frac{\partial \psi_{2}}{\partial  \phi_{1}^{T}}\right)^{T}$, $K_{1}= \textbf{K}_{rp}(\textbf{I}_{p}\otimes\boldsymbol\Gamma_{1})$ and $K_{2} = \textbf{K}_{rp}(\boldsymbol\eta^{T}\otimes\boldsymbol\Gamma_{0})$. Using straightforward matrix multiplication and corollary D1 to D3 in \cite{cook2010envelope} complete the proof.


\subsection{Proof of the comparison between the variance of the envelope and spatial envelope models}
\label{pro48}

In this section, we restate and proof the equation (4.8). 

For the simplify version of the spatial envelope and envelope, it can be shown that 
\begin{footnotesize}
\begin{eqnarray}
\label{compap}
\textbf{V}_{SPEN}^{-\frac{1}{2}}\textbf{V}_{EN}\textbf{V}_{SPEN}^{-\frac{1}{2}} =\frac{\textbf{X}^{T}\boldsymbol\rho^{-1}(\boldsymbol\theta)\textbf{X}}{n\sigma_{\textbf{X}}^{2}}
\textbf{I}_r +  \left( \frac{(\sigma_{0}^{2}-\sigma_{1}^{2})^{2}\left(1- \frac{\textbf{X}^{T}\boldsymbol\rho^{-1}(\boldsymbol\theta)\textbf{X}}{n\sigma_{\textbf{X}}^{2}}\right)}
{(\sigma_{0}^{2}-\sigma_{1}^{2})^{2}+\sigma_{1}^{2}\sigma_{\textbf{X}}^{2}||\boldsymbol\beta||^{2}}
\right)\boldsymbol\Gamma_{0}\boldsymbol\Gamma_{0}^{T},~
\end{eqnarray}
\end{footnotesize}
where $\textbf{V}_{SPEN}$ shows the asymptotic variance of the spatial envelope model, $\textbf{V}_{EN}$ shows the asymptotic variance of the envelope model, and $\sigma_{\textbf{X}}^{2}$ denotes the variance of the variance of the $\textbf{X}$ which is a $n\times1$ vector.

\textbf{Proof}: For the simplified version of the mode, the asymptotic variance for two models are:
\begin{eqnarray*}
var(\sqrt{n}\boldsymbol\beta_{Env}) &=& \frac{\sigma_{1}^{2}}{\sigma_{\textbf{X}}^{2}}\boldsymbol\Gamma_{1}\boldsymbol\Gamma_{1}^{T} +
\frac{\sigma_{0}^{2}\sigma_{1}^{2}\boldsymbol\eta^{T}\boldsymbol\eta}{\sigma_{\textbf{X}}^{2}\sigma_{1}^{2}\boldsymbol\eta^{T}\boldsymbol\eta+(\sigma_{0}^{2}-\sigma_{1}^{2})^{2}}
\boldsymbol\Gamma_{0}\boldsymbol\Gamma_{0}^{T}, \cr
var(\sqrt{n}\boldsymbol\beta^{*}) &=& \frac{n\sigma_{1}^{2}}{\textbf{X}^{T}\boldsymbol\rho^{-1}(\boldsymbol\theta)\textbf{X}}\boldsymbol\Gamma_{1}\boldsymbol\Gamma_{1}^{T} +
\frac{n\sigma_{0}^{2}\sigma_{1}^{2}\boldsymbol\eta^{T}\boldsymbol\eta}{\textbf{X}^{T}\boldsymbol\rho^{-1}(\boldsymbol\theta)\textbf{X}\sigma_{1}^{2} 
\boldsymbol\eta^{T} \boldsymbol\eta+n(\sigma_{0}^{2}-\sigma_{1}^{2})^{2}}\boldsymbol\Gamma_{0}\boldsymbol\Gamma_{0}^{T},~~
\end{eqnarray*}
therefore, to compare the variance of two models, we have 
\begin{footnotesize}
\begin{eqnarray*}
\textbf{V}_{SPEN}^{-\frac{1}{2}}\textbf{V}_{EN}\textbf{V}_{SPEN}^{-\frac{1}{2}} &=& \frac{\textbf{X}^{T}\boldsymbol\rho^{-1}(\boldsymbol\theta)\textbf{X}}{n\sigma_{\textbf{X}}^{2}}
\boldsymbol\Gamma_{1}\boldsymbol\Gamma_{1}^{T} + \frac{n(\sigma_{0}^{2}-\sigma_{1}^{2})^{2}+\sigma_{1}^{2}\textbf{X}^{T}\boldsymbol\rho^{-1}(\boldsymbol\theta)\textbf{X}\boldsymbol\eta^{T}\boldsymbol\eta}
{n(\sigma_{0}^{2}-\sigma_{1}^{2})^{2}+n\sigma_{1}^{2}\sigma_{\textbf{X}}^{2}\boldsymbol\eta^{T}\boldsymbol\eta}\boldsymbol\Gamma_{0}\boldsymbol\Gamma_{0}^{T} \cr
&=& \frac{\textbf{X}^{T}\boldsymbol\rho^{-1}(\boldsymbol\theta)\textbf{X}}{n\sigma_{\textbf{X}}^{2}}
\boldsymbol\Gamma_{1}\boldsymbol\Gamma_{1}^{T} + \frac{n(\sigma_{0}^{2}-\sigma_{1}^{2})^{2}+\sigma_{1}^{2}\textbf{X}^{T}\boldsymbol\rho^{-1}(\boldsymbol\theta)\textbf{X}\boldsymbol\eta^{T}\boldsymbol\eta}
{n(\sigma_{0}^{2}-\sigma_{1}^{2})^{2}+n\sigma_{1}^{2}\sigma_{\textbf{X}}^{2}\boldsymbol\eta^{T}\boldsymbol\eta}\boldsymbol\Gamma_{0}\boldsymbol\Gamma_{0}^{T} \cr
&\pm& 
\frac{\textbf{X}^{T}\boldsymbol\rho^{-1}(\boldsymbol\theta)\textbf{X}}{n\sigma_{\textbf{X}}^{2}}\boldsymbol\Gamma_{0}\boldsymbol\Gamma_{0}^{T} \cr
&=& \frac{\textbf{X}^{T}\boldsymbol\rho^{-1}(\boldsymbol\theta)\textbf{X}}{n\sigma_{\textbf{X}}^{2}}\textbf{I}_r +
\left(-\frac{\textbf{X}^{T}\boldsymbol\rho^{-1}(\boldsymbol\theta)\textbf{X}}{n\sigma_{\textbf{X}}^{2}} + 
\frac{n(\sigma_{0}^{2}-\sigma_{1}^{2})^{2}+\sigma_{1}^{2} \textbf{X}^{T}\boldsymbol\rho^{-1}(\boldsymbol\theta)\textbf{X}\boldsymbol\eta^{T}\boldsymbol\eta}
{n(\sigma_{0}^{2}-\sigma_{1}^{2})^{2}+n\sigma_{1}^{2}\sigma_{\textbf{X}}^{2}\boldsymbol\eta^{T}\boldsymbol\eta}
 \right)\boldsymbol\Gamma_{0}\boldsymbol\Gamma_{0}^{T} \cr
&=& \frac{\textbf{X}^{T}\boldsymbol\rho^{-1}(\boldsymbol\theta)\textbf{X}}{n\sigma_{\textbf{X}}^{2}}\textbf{I}_r + \frac{\textbf{X}^{T}\boldsymbol\rho^{-1}(\boldsymbol\theta)\textbf{X}}{n\sigma_{\textbf{X}}^{2}}
\left( -1 + \frac{\frac{n(\sigma_{0}^{2}-\sigma_{1}^{2})^{2}}{\textbf{X}^{T}\boldsymbol\rho^{-1}(\boldsymbol\theta)\textbf{X}}+\sigma_{1}^{2}\boldsymbol\eta^{T}\boldsymbol\eta}
{\frac{(\sigma_{0}^{2}-\sigma_{1}^{2})^{2}}{\sigma_{\textbf{X}}^{2}}+\sigma_{1}^{2}\boldsymbol\eta^{T}\boldsymbol\eta}
\right)\boldsymbol\Gamma_{0}\boldsymbol\Gamma_{0}^{T} \cr
&=& \frac{\textbf{X}^{T}\boldsymbol\rho^{-1}(\boldsymbol\theta)\textbf{X}}{n\sigma_{\textbf{X}}^{2}}\textbf{I}_r + \frac{\textbf{X}^{T}\boldsymbol\rho^{-1}(\boldsymbol\theta)\textbf{X}}{n\sigma_{\textbf{X}}^{2}}
\left( -1 + 1 + \frac{(\sigma_{0}^{2}-\sigma_{1}^{2})^{2}\left(\frac{n}{\textbf{X}^{T}\boldsymbol\rho^{-1}(\boldsymbol\theta)\textbf{X}}-\frac{1}{\sigma_{\textbf{X}}^{2}} \right)}
{\frac{(\sigma_{0}^{2}-\sigma_{1}^{2})^{2}}{\sigma_{\textbf{X}}^{2}}+\sigma_{1}^{2}\boldsymbol\eta^{T}\boldsymbol\eta}
\right)\boldsymbol\Gamma_{0}\boldsymbol\Gamma_{0}^{T} \cr
&=& \frac{\textbf{X}^{T}\boldsymbol\rho^{-1}(\boldsymbol\theta)\textbf{X}}{n\sigma_{\textbf{X}}^{2}}\textbf{I}_r + \frac{\textbf{X}^{T}\boldsymbol\rho^{-1}(\boldsymbol\theta)\textbf{X}}{n\sigma_{\textbf{X}}^{2}}
\left( \frac{(\sigma_{0}^{2}-\sigma_{1}^{2})^{2}\left(\frac{n\sigma_{\textbf{X}}^{2}}{\textbf{X}^{T}\boldsymbol\rho^{-1}(\boldsymbol\theta)\textbf{X}}-1 \right)}
{(\sigma_{0}^{2}-\sigma_{1}^{2})^{2}+\sigma_{1}^{2}\sigma_{\textbf{X}}^{2}\boldsymbol\eta^{T}\boldsymbol\eta}
\right)\boldsymbol\Gamma_{0}\boldsymbol\Gamma_{0}^{T} 
\end{eqnarray*}
\end{footnotesize}
Since $\boldsymbol\eta^{T}\boldsymbol\eta=||\boldsymbol\eta||^{2}=||\boldsymbol\beta||^{2}$, therefore we have
\begin{eqnarray*}
\frac{\textbf{V}_{SPEN}^{-\frac{1}{2}}\textbf{V}_{EN}\textbf{V}_{SPEN}^{-\frac{1}{2}}}{\frac{\textbf{X}^{T}\boldsymbol\rho^{-1}(\boldsymbol\theta)\textbf{X}}{n\sigma_{\textbf{X}}^{2}}} =
\textbf{I}_r + \left( \frac{(\sigma_{0}^{2}-\sigma_{1}^{2})^{2}\left(\frac{n\sigma_{\textbf{X}}^{2}}{\textbf{X}^{T}\boldsymbol\rho^{-1}(\boldsymbol\theta)\textbf{X}}-1 \right)}
{(\sigma_{0}^{2}-\sigma_{1}^{2})^{2}+\sigma_{1}^{2}\sigma_{\textbf{X}}^{2}||\boldsymbol\beta||^{2}}
\right)\boldsymbol\Gamma_{0}\boldsymbol\Gamma_{0}^{T} 
\end{eqnarray*}

\subsection{Preliminary Analysis for the Real Data}
\label{pre}
In this section, we provide the estimated Moran’s autocorrelation coefficient  (also called Moran's I) and empirical variogram for the real data. Moran's I is an extension of the Pearson correlation and measures spatial autocorrelation in the data \citep{cliff1973spatial}. For a a vector of data $s$, Moran's I is 
\[
MI = \frac{n}{S_0} \frac{\sum_{i=1}^{n}\sum_{j=1}^{n} w_{ij}(x_i-\bar{x})(x_j-\bar{x}) }{\sum_{i=1}^{n} (x_i-\bar{x})^2},
\]
where $\bar{x}$ denotes the mean of the observation, $w_{ij}$ is the weight between observation $i$ and $j$, and $S_0$ is the sum of all weights i.e. $S_0 = \sum_{i=1}^{n}\sum_{j=1}^{n} w_{ij}$. The weights $w_{ij}$, are chosen to be the inverse of the distance between observation $i$ and $j$. Using Moran's I, one can test the existence of the spatial autocorrelation where the null hypothesis is that there is no correlation versus the alternative hypothesis of there exists the spatial statistics. Table 4 presents the results of Moran's I for all the variables in the study. Based on these results, we can reject the null hypothesis that there is zero spatial autocorrelation present in the data for each variable. 

\begin{table}[htp]
\begin{center}
\caption{ Moran's I for different variables in the study.}  
\begin{tabular}{|l|c|c|c|c|} 
 \hline 
Variable& observed&   expected&     sd&         p.value    \\
 \hline 
Ozone&0.4498559 & -0.003731343 &0.02014298 &0           \\
 \hline 
Carbon monoxide&0.08161912 &-0.003731343 &0.01918668 &8.650319e-06\\
 \hline 
Sulfur dioxide&0.2425074 & -0.003731343 &0.01981788 &0           \\
 \hline 
Lead&0.234758  & -0.003731343 &0.01924146& 0           \\
 \hline 
Nitrogen dioxide&0.4414368 & -0.003731343 &0.02013472& 0           \\
 \hline 
Nitrogen monoxide&0.1665705  &-0.003731343 &0.01911524& 0           \\
 \hline 
PM 2.5&0.2449143 & -0.003731343 &0.02014268& 0           \\
 \hline 
PM 10&0.4063382  &-0.003731343 &0.01967082 &0   \\
\hline 
\end{tabular}
\end{center}
\end{table}

\begin{figure}
\centering
\includegraphics[width=0.4\textwidth, height= 5cm]{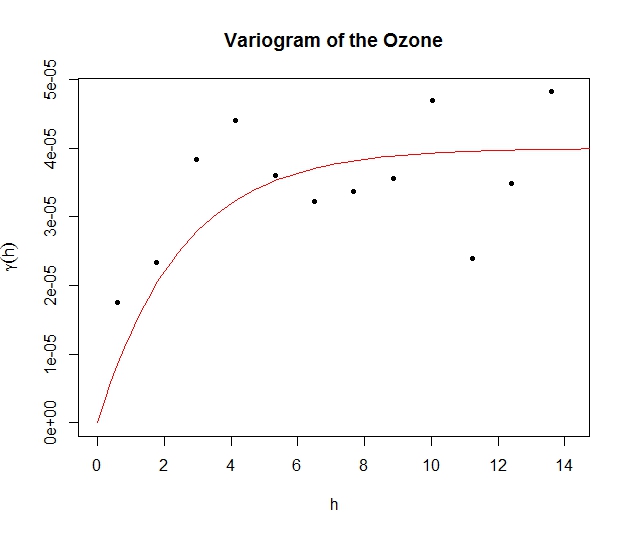}
\includegraphics[width=0.4\textwidth, height= 5cm]{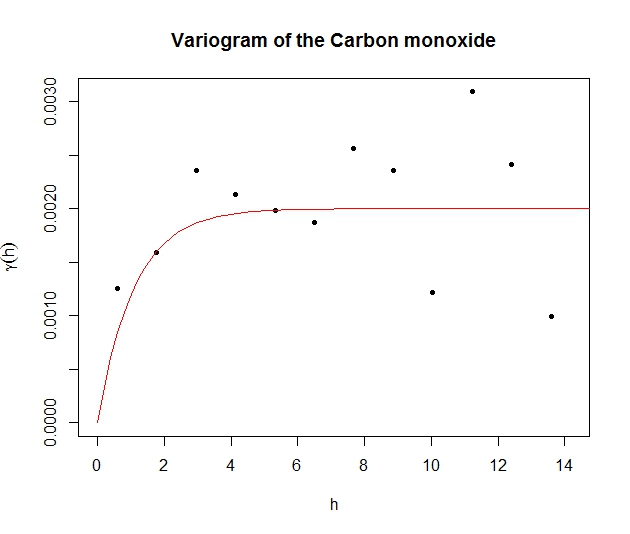}
\includegraphics[width=0.4\textwidth, height= 5cm]{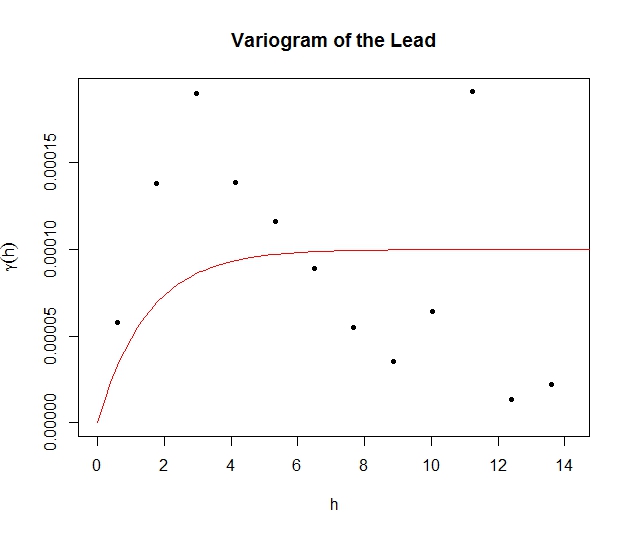}
\includegraphics[width=0.4\textwidth, height= 5cm]{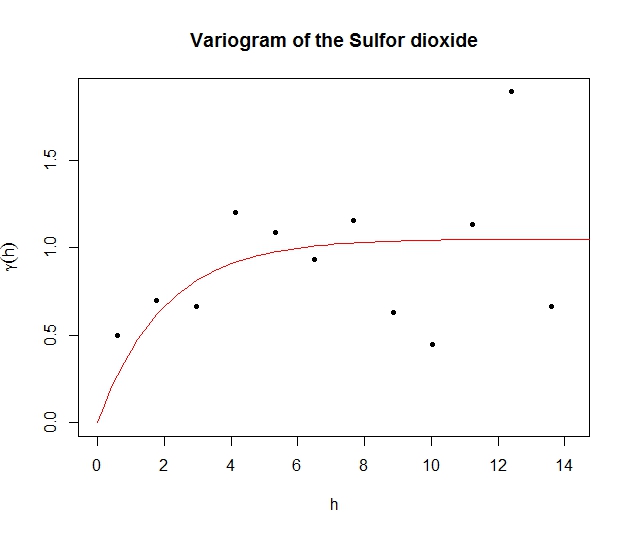}
\includegraphics[width=0.4\textwidth, height= 5cm]{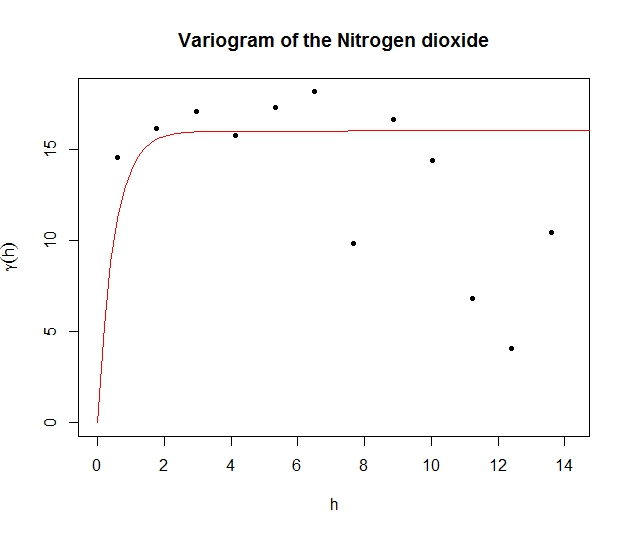}
\includegraphics[width=0.4\textwidth, height= 5cm]{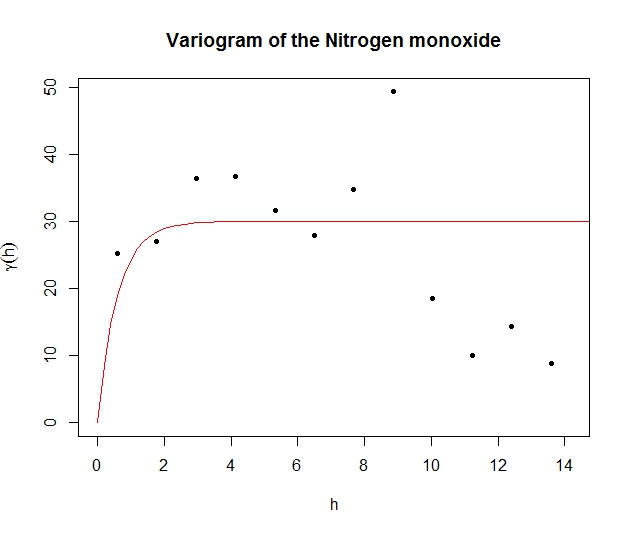}
\includegraphics[width=0.4\textwidth, height= 5cm]{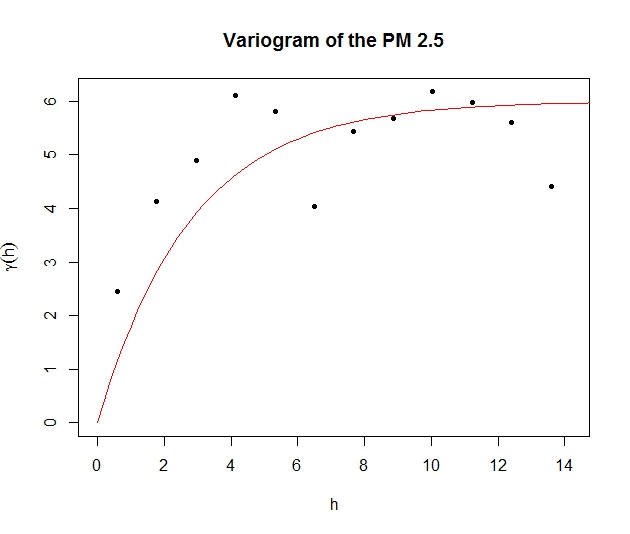}
\includegraphics[width=0.4\textwidth, height= 5cm]{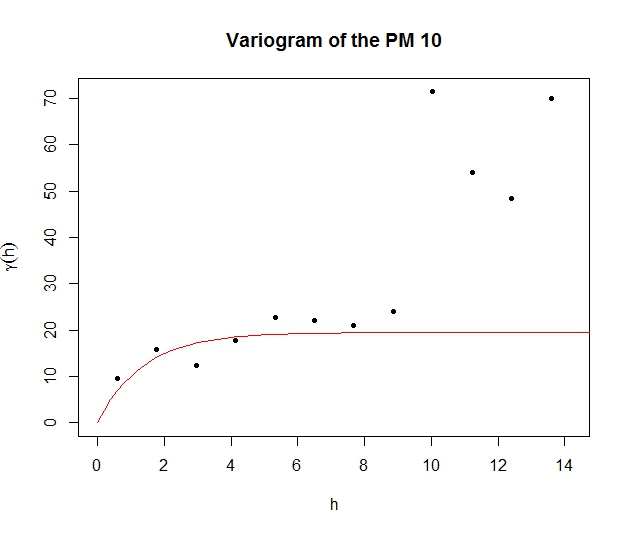}
\caption{The empirical variogram of  different responses in our study. These plots shows that using a Matern covariance function is reasonable.}
\label{figap4}
\end{figure}

In addition, to test the existence of the spatial correlation in the data, one common approach is to look at the patterns of the empirical variograms for the data in the  preliminary analysis. We used the Matern covariance function for the real data analysis. Using this covariance function makes the computation faster and it is one of the most common covariance function used in analyzing the air pollution data. Figure \ref{figap4} shows the empirical variogram of the responses. These plots show that using a Matern covariance function is reasonable.

\subsection{Estimated Regression Coefficients}
In this section, we provide the estimated regression coefficients and their standard deviation for traditional envelope model and our proposed model. As it can be seen the standard deviation for the estimated coefficients based on our proposed model is smaller than those calculated by traditional envelope model.

\begin{table}[htp]
\begin{center}
\caption{ Regression Coefficients (asymptotic standard deviation) using envelope the air pollution data in northeastern United States of America.}  
\begin{tabular}{|l|c|c|c|} 
 \hline 
Variable &  Relative humidity & Temperature &    Wind   \\   
 \hline 
Ozone &0.068 (0.388)&	-0.083 (0.493)&	-0.034 (0.303)	\\
\hline 
Carbon monoxide &-0.008 (0.051)&	0.014	(0.064)&	0.004	(0.040)	\\
\hline 
Lead &-0.016 (0.094)&	0.022	(0.120)&	0.008	(0.074)	\\
\hline 
Nitrogen dioxide &-0.050 (0.515)&	0.148	(0.564)&	0.037	(0.406)	\\
\hline 
Nitrogen monoxide &-0.032 (0.442)&	0.157	(0.553)&	0.001	(0.346)	\\
\hline 
Sulfur dioxide & -0.029 (0.381)&	0.196	(0.487)&	0.007	(0.297)	\\
\hline 
PM10 &0.013	(0.353)&	0.188	(0.440)&	-0.021 (0.276)	\\
\hline 
PM2.5   &0.033	(0.343)&	-0.162 (0.581)&	-0.011 (0.261)	\\
\hline 
\end{tabular}
\end{center}
\end{table}

\begin{table}[htp]
\begin{center}
\caption{ Regression coefficients (asymptotic standard deviation) using spatial envelope the air pollution data in northeastern United States of America.}  
\begin{tabular}{|l|c|c|c| }
 \hline 
Variable &  Relative humidity & Temperature &    Wind   \\   
 \hline 
Ozone & 0.007	(0.178)&	-0.004	(0.083)&	-0.004	(0.033)	\\
\hline 
Carbon monoxide &	 0.011	 (0.005)&	0.014	(0.064)&	-0.001	(0.001)	\\
\hline 
Lead &	-0.001	(0.014)&	0.002	(0.120)&	0.001	(0.004)	\\
\hline 
Nitrogen dioxide &	0.072	(0.021)&	0.348	(0.121)&	-0.037	(0.046)	\\
\hline 
Nitrogen monoxide &0.062	(0.022)&	0.457	(0.115)&	-0.084	(0.023)	\\
\hline 
Sulfur dioxide &	-0.613	(0.111)&	0.196	(0.006)&	0.004	(0.096)	\\
\hline 
PM10 &	-0.013	(0.025)&	0.188	(0.024)&	-0.098	(0.026)	\\
\hline 
PM2.5   &	0.116	(0.143)&	0.162	(0.051)&	0.003	(0.016)	\\
\hline 
\end{tabular}
\end{center}
\end{table}

\newpage


\subsection{Prediction Plot for Response Variables}

\begin{figure}[h]
    \centering
    \includegraphics[width=0.8\textwidth,height=6cm]{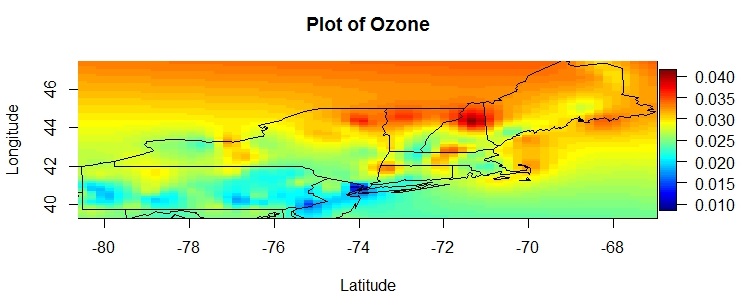}
    \caption{Prediction plot of the log of the ground level Ozone for the study area. As it can be seen, the Ozone level is not high in the study area. The north part of New Hampshire seems to have the highest value for the Ozone.}
    \label{fig4}
\end{figure}

\begin{figure}
    \centering
    \includegraphics[width=0.8\textwidth,height=6cm]{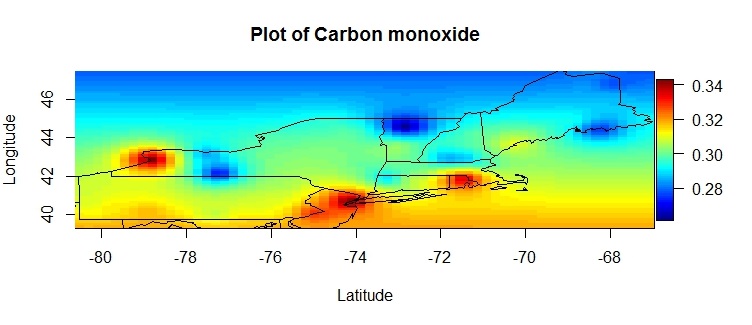}
    \caption{Prediction plot of carbon monoxide (CO) for the study area. As it can be seen, the carbon monoxide is moderately low in the study area. CO is high in Rhodes Island, New York, New Jersey, and Buffalo which are highly populated and therefore there will be a lots of car and usage of fossil fuels which leads to high concentration of carbon monoxide in the air.}
    \label{fig3}
\end{figure}

\begin{figure}
    \centering
    \includegraphics[width=0.8\textwidth,height=6cm]{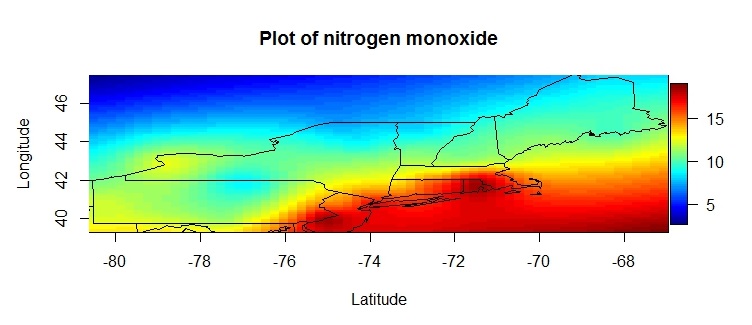}
    \caption{Prediction plot of the Nitrogen monoxide for the study area. as it can be seen, the Nitrogen monoxide is high in New York and New Jersey and moderately high almost every place in the study area. }
    \label{fig10}
\end{figure}

\begin{figure}
    \centering
    \includegraphics[width=0.8\textwidth,height=6cm]{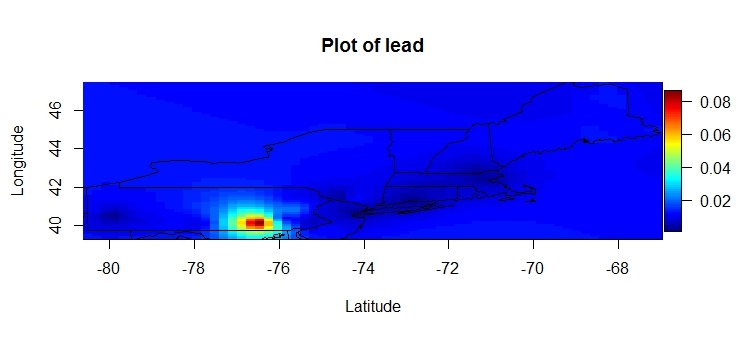}
    \caption{Prediction plot of lead for the study area. As it can be seen, the lead is high in Harrisburg and Lancaster.}
    \label{fig2}
\end{figure}

\end{document}